\documentstyle[aps,epsfig]{revtex} \def\narrowtext{} %%\tighten %\twocolumn
\begin{document} 
\newcommand{\dprime}{{\prime\prime}}
\newcommand{\be}{\begin{equation}}
\newcommand{\den}{\overline{n}} 
\newcommand{\ee}{\end{equation}}
\newcommand{\bea}{\begin{eqnarray}} 
\newcommand{\eea}{\end{eqnarray}}
\newcommand{\nn}{\nonumber} 
\newcommand{\vk}{{\bf k}}
\newcommand{\vv}{{\bf v}} 
\newcommand{\la}{\langle}
\newcommand{\ra}{\rangle} 
\newcommand{\ph}{\phi} 
\newcommand{\dg}{\dagger}
\renewcommand{\vr}{{\bf{r}}} 
\newcommand{\vq}{{\bf{q}}}
\newcommand{\vQ}{{\bf{Q}}} 
\newcommand{\hj}{\hat{\alpha}}
\newcommand{\hx}{\hat{\bf x}} 
\newcommand{\hy}{\hat{\bf y}}
\newcommand{\hz}{\hat{\bf z}}
\newcommand{\vS}{{\bf S}} 
\newcommand{\cV}{{\cal U}}
\newcommand{\cD}{{\cal D}} 
\newcommand{\tnh}{{\rm tanh}}
\newcommand{\sh}{{\rm sech}} 
\newcommand{\vR}{{\bf R}}
\newcommand{\crx}{c^\dg(\vr)c(\vr+\hx)}
\newcommand{\crkubox}{c^\dg(\vr)c(\vr+\hat{x})}
\newcommand{\pll}{\parallel} 
\newcommand{\crj}{c^\dg(\vr)c(\vr+\hj)}
\newcommand{\crmj}{c^\dg(\vr)c(\vr - \hj)}
\newcommand{\sumall}{\sum_{\vr}} 
\newcommand{\sumx}{\sum_{r_1}}
\newcommand{\nabj}{\nabla_\alpha \theta(\vr)} 
\newcommand{\nabx}{\nabla_1\theta(\vr)} 
\newcommand{\sumy}{\sum_{r_2,\ldots,r_d}}
\newcommand{\krj}{K(\vr,\vr+\hj)} 
\newcommand{\sigr}{|\psi_0\rangle}
\newcommand{\sigl}{\langle\psi_0 |}
\newcommand{\sier}{|\psi_{\Phi}\rangle}
\newcommand{\siel}{\langle\psi_{\Phi}|}
\newcommand{\sumrj}{\sum_{\vr,\alpha=1\ldots d}}
\newcommand{\krw}{K(\vr,\vr+\hx)} 
\newcommand{\Dtheta}{\Delta\theta}
\newcommand{\rhonew}{\hat{\rho}(\Phi)}
\newcommand{\rhoold}{\hat{\rho_0}(\Phi)} 
\newcommand{\dt}{\delta\tau}
\newcommand{\cP}{{\cal P}} 
\newcommand{\cS}{{\cal S}}
\newcommand{\vm}{{\bf m}} 
\newcommand{\hnr}{\hat{n}({\vr})}
\newcommand{\hnm}{\hat{n}({\vm})} 
\newcommand{\del}{\hat{\delta}}
\newcommand{\upa}{\uparrow} 
\newcommand{\dna}{\downarrow}
%\begin{multicols}{1}

\draft

\title{
Effective actions and phase fluctuations in $d$-wave superconductors
}
\author {Arun Paramekanti$^{1}$, Mohit Randeria$^{1}$, T. V. Ramakrishnan$^{2}$
and S.S. Mandal $^{2,3}$.
       }
\address{
         $^{1}$Tata Institute of Fundamental Research, Mumbai 400 005, India \\
         $^{2}$Indian Institute of Science, Bangalore 560 012, India \\
         $^{3}$J.N. Centre for Advanced Scientific Research, Bangalore 560 094, India
         }

\address{%
\begin{minipage}[t]{6.0in}
\begin{abstract}
We study effective actions for order parameter fluctuations
at low temperature in layered $d$-wave superconductors such as the cuprates.
The order parameter lives on the bonds of a square lattice and 
has two amplitude and two phase modes associated with it.
The low frequency spectral weights for amplitude and relative phase
fluctuations is determined and found to be subdominant to quasiparticle
contributions. The Goldstone phase mode and 
its coupling to density fluctuations
in charged systems is treated in a gauge-invariant manner. The
Gaussian phase action is used to study both the $c$-axis Josephson plasmon 
and the more conventional in-plane plasmon in the cuprates. 
We go beyond the Gaussian theory by 
deriving a coarse-grained quantum XY model, which incorporates important
cutoff effects overlooked in previous studies.
A variational analysis of this effective model shows that in the 
cuprates, quantum effects of phase fluctuations are 
important in reducing the zero temperature superfluid stiffness, but thermal 
effects are small for $T \ll T_c$.
\typeout{polish abstract}
\end{abstract}
\end{minipage}}

\maketitle
\narrowtext

\noindent

\section{Introduction} 

The high temperature cuprate superconductors (SCs) 
differ from conventional SCs 
in several respects: a $d$-wave gap with gapless quasiparticle 
excitations, a small superfluid phase 
stiffness, a short coherence length and strong electron interactions. It 
is therefore of interest to examine some of these unconventional aspects 
and their interplay in simple 
models of high $T_c$ systems. With this motivation,
we study in this paper, the low temperature collective properties of charged, 
layered 
$d$-wave SCs with a short coherence length and small superfluid stiffness.

The superfluid phase stiffness, $D_s = n_s/m^*$, is a fundamental property 
characterizing SCs \cite{scalapino93}, which is directly related to 
$\lambda$, the experimentally measured magnetic penetration depth 
\cite{lambda} in the London limit .
The low temperature behavior of $D_s$ contains information
on the low-lying excitations in these systems. 
Experimentally, $\lambda(T)$ is found to increase linearly with $T$ in
the high $T_c$ SCs \cite{lambdarefs}, implying a linearly decreasing
$D_s$. This linear drop in $D_s$ has been attributed to
quasiparticle excitations near the nodes of the $d$-wave gap.
Alternatively, it has been suggested that this effect could arise
entirely from classical thermal phase fluctuations \cite{roddick95,carlson99}
and quasiparticles can be ignored \cite{roddick95,carlson99}. 
It is then clearly of interest to 
identify the important low energy 
excitations in these systems, from the point of view of understanding
the penetration depth data, as well as other thermodynamic properties
and response functions.

From a theoretical perspective, the physics of a system with a small 
superfluid stiffness and short coherence length has been studied in 
detail in case of neutral $s$-wave SCs \cite{BECreview,JRE97}.
In this case, the fermionic excitations and 
fluctuations in the order parameter
amplitude are gapped, and phase fluctuations are the only important 
excitation at low temperature. It is of interest to compare this
with the behavior in models which support an anisotropic order parameter 
with low lying fermionic excitations, such as a $d$-wave SC.

We approach the problem by deriving and analyzing effective actions 
for a $d$-wave SC within a functional integral framework, which
allows us to focus on the collective (order parameter) degrees of freedom.
We note that our effective actions are derived by looking at fluctuations
around a BCS mean field solution. We believe that such an approach
is valid for the SC state of the high Tc materials, at least for $T \ll T_c$.
There is considerable experimental evidence for sharply defined
quasiparticle excitations about the d-wave SC state, and thus the ground 
state and low-lying excitations appear to be adiabatically connected to their
BCS counterparts. We thus expect that while strong correlations will
modify the {\em coefficients} of the phase action, they will not change its 
qualitative form.

Our main results can be summarized as follows:

1. We find that the $d$-wave SC state is characterized in
terms of an order parameter which lives on the bonds of a square lattice.  
The bond order parameter leads to two amplitude and two phase modes in
contrast to $s$-wave SCs. One of the phase modes is identified as the 
usual (Goldstone) phase mode while the other, which we call
the ``bond phase'', is the relative phase between the $x$ and $y$ bonds 
at a site. The latter can be thought of as representing 
fluctuations from the $d$-wave state towards an extended $s$-wave state. The
amplitude and bond-phase fields have spin zero and
couple to the particle-particle channel. 

2. We study the spectral weight
for fluctuations of the amplitude and bond-phase fields and 
find that they are {\em not gapped} but rather exhibit power laws
down to zero energy. However, the low energy spectral weight in these 
fluctuations is very small compared to the quasiparticle contribution.

3. We derive an effective Gaussian action for the usual phase variable in 
charged systems, since this couples directly to the electromagnetic 
potentials. The large in-plane plasma frequency, which is relatively
unaffected by superconductivity, and 
the low energy $c-$axis Josephson plasmon at zero and finite temperatures 
are studied in a unified manner within the same formalism. We emphasize the
relation between unusual aspects of the c-axis optical conductivity 
and the Josephson plasmon. We also discuss the plasmon
dispersion in layered systems.

4. We extend the above formalism to consider the effect of phase
fluctuations beyond Gaussian level on the superfluid stiffness
in charged systems. The quantum XY model \cite{quantumXY} is
usually used for such an analysis, motivated by studies of Josephson
junction arrays and granular SCs. We emphasize that there are important
differences when considering low temperature bulk SCs, and derive
a quantum XY phase action suitable for our problem, correctly taking
into account appropriate momentum and frequency cutoffs, missed
in earlier studies. 

5. The low temperature renormalization of $D_s$ by 
phase fluctuations is studied within a self-consistent harmonic 
approximation. 
For parameter values relevant to the cuprates near optimal doping,
quantum phase fluctuations are shown to lead to a sizeable renormalization 
of the superfluid stiffness. However, thermal fluctuations are found
to have no effect at low temperatures, 
unlike the results of earlier
studies \cite{roddick95,carlson99}. These studies focussed on the effect
of thermal phase fluctuations, but Coulomb effects were considered to be 
unimportant, in contrast to the present work.

6. As part of our analysis, we also touch upon certain formal
issues which may be of some general interest. Among these are:
(a) how gauge invariance can be understood in a simple manner within the 
functional integral language; (b) the role of the linear time derivative 
term $i\rho\partial\tau\theta$ in the phase action; and 
(c) the problems involved in deriving local phase actions which respect
$2\pi$ periodicity starting from a fermionic model.

The paper is organized as follows. In Section II, we present the
Hamiltonian for our model, and discuss the effective action and mean
field theory in Section III. Section IV contains a discussion of
fluctuations of the amplitude and bond-phase fields with some of the details
discussed in Appendix A. In Section V, we turn to phase fluctuations
and derive effective phase-only actions for neutral and charged systems. 
The linear time derivative term in the action which arises in this context 
is briefly discussed in Appendix B. We then derive
gauge invariant density and current correlations, leaving details of the
algebra to Appendix C. In Section VI we discuss collective in-plane and 
$c-$axis 
plasmons. In Section VII we present the derivation of a quantum XY
model appropriate for charged, layered SCs. We analyze this action
and compute the renormalization of the phase stiffness by longitudinal phase 
fluctuations in Section VIII and discuss experimental
implications. We conclude in Section IX with a discussion and summary of 
our results.

\section{The Hamiltonian}

We consider a system of fermions with kinetic energy
$K = \sum_{k,\sigma} \xi_\vk c^{\dg}_{\vk,\sigma} c_{\vk,\sigma}$
(where $\xi_\vk = \epsilon_\vk - \mu$ with $\epsilon_\vk$ the
2D dispersion and $\mu$ the chemical potential) interacting via
a separable potential which is attractive in the $d$-wave channel. 
We will show that in coordinate space this interaction leads to the 
superexchange term of the t-J model.

Let us begin with 
\be
H'_{\rm pair} =
- \frac{g}{N}\sum_{ \vk,\vk',\vq } \varphi_d(\vk)\varphi_d(\vk')
c^\dg_{\vk+\vq/2 \upa} c^\dg_{-\vk+\vq/2 \dna}
c^{ }_{-\vk'+\vq/2 \dna} c^{ }_{\vk'+\vq/2 \upa}
\label{bcsk}
\ee
where $\varphi_d(\vk)=(\cos k_x - \cos k_y)$.
and we work on a 2D square lattice with lattice spacing $a$.
$\Omega$ denotes the volume of the system with $N$ sites.
We set $a=1$ in most equations, but retain it in some for the sake of 
clarity.
Fourier transforming to real space, we get
\be
H_{\rm pair} =
-\frac{g}{4} \sum_{\la \vr,\vr'\ra} B^\dg_{\vr,\vr'}B_{\vr,\vr'} 
= \frac{g}{2}\sum_{\la \vr,\vr'\ra} ({\bf S}_\vr\cdot
{\bf S}_{\vr'}- \frac{1}{4}n_\vr n_{\vr'}).
\label{bcsr}
 \ee
(The prime on $H_{\rm pair}$ is omitted in going from (\ref{bcsk}) to
(\ref{bcsr}) for reasons explained below).
Here $\la \vr,\vr' \ra$ are nearest neighbour sites, and
$B^\dg_{\vr, \vr'} \equiv c^\dg_{\vr,\upa} c^\dg_{\vr',\dna}
 - c^\dg_{\vr,\dna} c^\dg_{\vr',\upa}$
creates a singlet on the bond $(\vr,\vr')$, while
${\bf S}_\vr$ and $n_\vr$ are the spin and number operators.
Clearly this is the interaction term of the $t-J$
model with $g = 2J$.

There is a subtlety involved here; on transforming (\ref{bcsr})
back to $\vk-$space we do {\it not} recover the original
expression (\ref{bcsk}) we had started with. Instead we obtain,
using from now on $g = 2J$,
\be
H_{\rm pair} = - \frac{J}{N}\sum_{\vk,\vk',\vq} 
\left[\varphi_d(\vk) \varphi_d(\vk') 
+ \varphi_{s}(\vk) \varphi_{s}(\vk')\right]
c^\dg_{\vk+\vq/2 \upa} c^\dg_{-\vk+\vq/2 \dna}
c_{-\vk'+\vq/2 \dna} c_{\vk'+\vq/2 \upa}
\label{ssdd}
\ee
where
$\varphi_{s}(\vk)=(\cos k_x +\cos k_y)$.
The reason for two different ${\bf k}$-space interactions
leading to the same real space expression is the following operator
identity on a 2D square lattice:
\be
\sum_{\vk,\vk',\vq} 
\left[\varphi_d(\vk) \varphi_d(\vk')
- \varphi_{s}(\vk) \varphi_{s}(\vk')\right]
c^\dg_{\vk+\vq/2 \upa} c^\dg_{-\vk+\vq/2 \dna}
c_{-\vk'+\vq/2 \dna} c_{\vk'+\vq/2 \upa}
\equiv 0
\label{idntty}
\ee

It can be shown that (\ref{bcsr}) and (\ref{ssdd}) 
both lead to the same self consistent BCS gap equation.
Thus we will use the interaction in (\ref{ssdd}), and not (\ref{bcsk}).
From the form of (\ref{ssdd}), it is clear that $H_{pair}$ has attraction
in both the $d$-wave channel, with a $\varphi_d(\vk)$ order parameter,
and in the extended $s$-wave ($s^*$) channel, with a 
$\varphi_{s}(\vk)$ order parameter. 

We will now analyze the Hamiltonian $H = K + H_{\rm pair}$.
Later, we will also add to it the Coulomb interaction 
appropriate to layered systems (see Section V.B). 

\section{Mean Field Theory}

The partition function at a temperature $T$ is written as the standard 
coherent state path integral 
with the action
$
\int_0^{1/T}~d\tau \left[\sum_{\vr,\sigma} c^\dg_{\vr, \sigma}
\partial_\tau c_{\vr, \sigma}
+ H \left(c,c^\dg\right)\right]$.
We decouple $H_{\rm pair}$ with a complex 
field $\Delta_{\vr,\vr'}(\tau)$ 
using the Hubbard-Stratonovich transformation:
\be
\exp\left(\frac{J}{2} 
B^\dg_{\vr,\vr'}(\tau) B_{\vr,\vr'}(\tau)\right)
= \int ~\cD\left(\Delta~\Delta^*\right)
 \exp\left[-{\cal L}(\vr,\vr';\tau)\right]
\label{decoupD1}
\ee
where
\be
{\cal L}(\vr,\vr';\tau) = \frac{1}{8 J} |\Delta_{\vr,\vr'}(\tau)|^2 
- \frac{1}{4} \left( \Delta_{\vr,\vr'}(\tau) B^\dg_{\vr,\vr'}(\tau) +
  {\rm h.c.} \right).
\label{decoupD2}
\ee
We thus obtain 
the action
\be
S = \int_0^{1/T}d\tau\left[ \sum_{\vk,\sigma} c^{\dg}_{\vk,\sigma}(\tau)
\left(\partial_\tau + \xi_\vk \right) c_{\vk,\sigma}(\tau)
+ \sum_{\la\vr,\vr'\ra} {\cal L}(\vr,\vr';\tau)
\right].
\label{basicaction}
\ee
The fermion fields
can then be integrated out to obtain the effective action
$\exp\left(-S_{\rm eff}\left[\Delta,\Delta^*\right]\right)
= \int \cD\left(c,c^\dg\right) e^{-S}.$

The $d$-wave saddle point is given by
$\Delta_{\vr,\vr+\hx}(\tau) = -\Delta_{\vr,\vr+\hy}(\tau) = \Delta_d$, 
a $(\vr,\tau)$-independent real number, obtained from 
$\delta S_{\rm eff}/\delta \Delta_d = 0$, which leads to
the BCS gap equation 
\be
\frac{1}{J} = \frac{1}{N}\sum_\vk \frac{\varphi^2_d(\vk)}
{2 E_\vk}~
{\rm tanh}(E_\vk/2T)
\label{gapeqn}
\ee
where $E_\vk=\sqrt{\xi^2_\vk+\Delta^2_\vk}$ and $\Delta_\vk
=\Delta_d \varphi_d(\vk)/2$.
The same gap equation can be obtained by starting from the 
momentum space potential in (\ref{ssdd}) and decoupling in the 
$d$-wave channel.

Given $H_{pair}$ of (\ref{ssdd}) one could equally
well look for possible extended $s$-wave ($s^*$) saddle points with 
$\Delta_{\vr,\vr+\hx}(\tau)=\Delta_{\vr,\vr+\hy}(\tau)=\Delta_{s}$. 
However, for our choice of dispersion \cite{param} 
(which includes nearest and
next nearest neighbour hopping with opposite signs)
we have found by numerical solution of the gap equation that the 
$d$-wave saddle point is stable relative to the $s^*$ solution.
The reason for this can be seen as follows:
$\varphi_{s}(\vk)$ is small over most of the Fermi surface for the
fillings of interest, while $\varphi_d(\vk)$
vanishes only on the nodes. Thus the condensation energy gained by the
$s^*$ state is smaller than the $d$-wave state. Further, if we consider
the large on-site repulsion between electrons (which we have
not done here, but is certainly an essential part of the full
$t-J$ model) and demand
$\la c^\dg_{\vr,\upa} c^\dg_{\vr,\dna} \ra = 0$, then we have $\sum_\vk
\Delta_\vk/2 E_\vk = 0$. This is automatically satisfied for the
$d$-wave state at any filling, but not for the $s^*$ state. In this work, 
we rely on the former ``Fermi surface effect'' to stabilize the $d$-wave 
state.

\section{Fluctuations}

To treat fluctuations in the order parameter we
write $\Delta_{\vr,\vr'}(\tau) = |\Delta_{\vr,\vr'}(\tau)| 
e^{i \Phi_{\vr,\vr'}(\tau)}$.
The phase
$\Phi_{\vr,\vr+\hx}(\tau) = 0$ and $\Phi_{\vr,\vr+\hy }(\tau) = \pi$ 
at the $d$-wave saddle point. 
We now divide the phase field into two parts;
following Ref\cite{balents98}, we set 
$\Phi_{\vr,\vr+\hat{x}}(\tau)=\theta(\vr,\tau)$ and 
$\Phi_{\vr,\vr+\hat{y}}(\tau)=\pi+\phi(\vr,\tau)+\theta(\vr,\tau)$. 

We next assume that the spatial variation of $\theta(\vr,\tau)$
is small on the scale of the lattice spacing, which allows 
us to set 
$\Phi_{\vr,\vr+\hat{x}}(\tau) \approx \frac{1}{2}\left[\theta(\vr,\tau)+\theta(
\vr+\hat{x},\tau)\right]$ and 
$\Phi_{\vr,\vr+\hat{y}}(\tau) \approx \pi+\phi(\vr,\tau)+
\frac{1}{2}\left[\theta(\vr,\tau)+\theta(\vr+\hat{y},\tau)\right]$.
While we lose the 
$\theta(\vr,\tau)\to \theta(\vr,\tau) +2\pi$ 
invariance of the action with this approximation,
it is nevertheless useful in isolating
that part of the phase field which couples to electromagnetic 
fields as we will see below. 
We can now transform to new fermion variables given by $c^\dg_\vr(\tau)
\to
c^\dg_\vr(\tau) e^{-i \theta(\vr,\tau)/2}$ \cite{ambegaokar_NATO,tvr89}.
As a result of this ``gauge transformation'' the action
of (\ref{basicaction}) gets modified to
\be
S = \int_0^{1/T}d\tau\left[
{\cal L}_0 + {\cal L}_1 \right]
\label{fullaction}
\ee
with
\be
{\cal L}_0 =
\sum_{\vr,\sigma} c^\dg_{\vr,\sigma}(\tau) e^{-i\theta(\vr,\tau)/2}
(\partial_\tau-\mu)c_{\vr,\sigma}(\tau)e^{i\theta(\vr,\tau)/2}
-\frac{1}{2} \sum_{\vr,\vr',\sigma} t(\vr-\vr')
\left[ c^\dg_{\vr,\sigma}(\tau)c_{\vr',\sigma}(\tau)
e^{-i[\theta(\vr,\tau)-\theta(\vr',\tau)]/2} + {\rm h.c.}\right]
\label{lag0}
\ee
where $t(\vr-\vr')$ is the hopping matrix element between 
point $\vr$ and $\vr'$, so that
$\epsilon_\vk = \sum_{\vr-\vr'} e^{i\vk\cdot (\vr-\vr')}t(\vr-\vr')$ and
\be
{\cal L}_1 =
\frac{1}{8 J} \sum_{\la \vr,\vr' \ra} |\Delta_{\vr,\vr'}(\tau)|^2 
- \frac{1}{4} \sum_{\vr} |\Delta_{\vr,\vr+\hx}(\tau)|
  \left(B^\dg_{\vr,\vr+\hx}(\tau) + {\rm h.c.} \right)
+ \frac{1}{4} \sum_{\vr} |\Delta_{\vr,\vr+\hy}(\tau)|
  \left(B^\dg_{\vr,\vr+\hy}(\tau) e^{i\phi(\vr,\tau)}
   + {\rm h.c.} \right).
\label{lag1}
\ee
In the following Sections we shall integrate out the fermions and
examine the resulting effective actions 
for the amplitude, $\phi$ and $\theta$ fields.

\subsection{Amplitude fluctuations}

Amplitude fluctuations can be considered by setting
$|\Delta_{\vr,\vr+\hat{\alpha}}(\tau)|=\Delta_d(1+\eta_\alpha(\vr,\tau))$
in (\ref{lag1}), where $\alpha=x,y$.  The transformation from 
$\Delta_\alpha,\Delta_\alpha^*$ to $\eta_\alpha,\theta,\phi$ has a 
Jacobian $4 \Delta^4_d (1+\eta_x) (1+\eta_y)$ at every
point $(\vr,\tau)$, leading to an additional term
\be
\int_0^{1/T}d\tau~{\cal L}_2 = T \int_0^{1/T}d\tau~\sum_{\vr,\alpha} 
\ln \left(1+\eta_\alpha(\vr,\tau)\right)
\approx T \int_0^{1/T}d\tau~\sum_{\vr,\alpha} \left[ \eta_\alpha(\vr,\tau) - 
\frac{1}{2}\eta_\alpha^2(\vr,\tau) \right] 
\label{lag2}
\ee
in the action in (\ref{fullaction}). 
For $(\vq,\omega)\neq(0,0)$, the linear 
term in (\ref{lag2}) can be set to zero and only the quadratic term 
contributes. However, even this term is zero at $T=0$ and can be ignored at 
low $T$. 

From (\ref{lag1}) we see that the spin zero amplitude fields 
$\eta_{\alpha}$ couple to singlet pairs.
Their coupling to the $\phi$-field 
can be shown to be small at small momentum and frequency.
In particular, for static uniform distortions of $\phi$ and
$\eta_{_\alpha}$ the energy has to be even under $\phi\to -\phi$ and terms
like $\phi \eta_{_\alpha}$ cannot appear in the action
on integrating out the fermions. 
The mixing of $\eta_{\alpha}$
with the phase $\theta$ and electromagnetic potentials can also be shown to 
be negligible at small $\vq,\omega$ 
since $\eta_{\alpha}$ couples to the particle-particle channel while the 
$\theta$ and electromagnetic potentials couple to the particle-hole channel. 
The mixing then involves integrals over products of ordinary and 
anomalous Green functions
which vanish using particle-hole symmetry near the Fermi surface and the
$\vk-$dependence of $\Delta_\vk$. 
This lack of mixing of amplitude and phase modes
is similar to the well known weak coupling result for 
$s$-wave superconductors \cite{JRE97}.

Unlike the $s$-wave case, however, the amplitude excitations have
low frequency spectral weight in $d$-wave SCs which we now estimate.
Setting the phase fields to saddle point values and integrating out 
the fermions we obtain an effective action for amplitude fluctuations.
Transforming to new variables $\eta_{s}=(\eta_x+\eta_y)/\sqrt{2}$ and
$\eta_d=(\eta_x-\eta_y)/\sqrt{2}$,
diagonalizes the action for $\eta$ fields at $\vq=0$ and is
a good starting point to consider small $\vq$ fluctuations.
We obtain, to Gaussian order,  $S[\eta] = \frac{1}{2 T} \sum_{\vq,n,i=s,d} 
\eta^*_i M^{-1}_i(\vq,i\omega_n) \eta_i$, where $\omega_n=2\pi n T$.
We have made the approximation of
ignoring the coupling between $\eta_s$ and $\eta_d$, which can be shown
to be negligible at small momentum.

The low energy density of states is given by $N_i (\omega) = 
\frac{1}{N}\sum_\vq {\cal I}m M_i(\vq,\omega)/\pi$ with 
$|\vq_x|,|\vq_y| < \pi/\xi_0$. The cutoffs arise
since the fluctuations must have an energy lower than
the condensation energy as discussed in more detail later, in
the context of phase fluctuations. 
From a numerical calculation of $N_{s,d}(\omega)$ we find
that $N_s(\omega)/N_{qp}(\omega) \sim \omega^2/(\Delta_d v_{_F}/a)$ and
$N_d(\omega)/N_{qp}(\omega) \sim \omega^4/(\Delta_d^3 v_{_F}/a)$ where
$N_{qp}(\omega)\equiv k_{_F}\omega/(\pi v_{_F}\Delta_d)$ is the
density of states per spin for quasiparticle excitations. 
These results can also be understood from an approximate analysis of the
form of $M_{s,d}(\vq,\omega)$ discussed in Appendix A.

We see that both $N_s(\omega)$ and $N_d(\omega)$ 
are much smaller than $N_{qp}(\omega)$ for $\omega \ll \Delta_d$,and
thus conclude that amplitude fluctuations
are unimportant for low temperature properties which will be dominated
by the quasiparticle contribution.

\subsection{The ``bond-phase'' field $\phi$}

We next study the field $\phi$.
We note from (\ref{lag1}) 
that a uniform $\phi=\pi$ would lead to extended $s$-wave ($s^*$)
order. One can therefore think of $\phi$ as representing fluctuations of 
$s^*$ character about the $d$-wave saddle point. 
From (\ref{lag1}) we see $\phi$ is a spin zero field which couples to pairs.

For reasons similar to those explained above for amplitude fluctuations,
the coupling of $\phi$ to other fields is weak and may be ignored at 
low momentum and frequency.
We therefore derive a Gaussian action for $\phi$
by setting $\theta = \eta_{_\delta} = 0$, their saddle point values,
and integrating out the fermions. 
This leads to the action
$S[\phi]
= \frac{1}{T} \sum_{n,\vq} M^{-1}_\phi(\vq,\omega_n) |\phi(\vq,\omega_n)|^2$.
Since the $\phi$ field has low frequency spectral weight,
we compute its density of states
$ N_\phi(\omega) = \frac{1}{N \pi}
\sum_{|\vq|<\pi/\xi_0}{\cal I}m M_\phi(\vq,\omega+i 0^+)$.
From numerical calculations, as well as simpler
approximations discussed in Appendix A, we find 
$N_\phi(\omega)/N_{qp}(\omega) \sim \omega^2/(\Delta_d v_{_F}/a)$.
We thus see that $\phi$ fluctuations are much less important than
quasiparticles at low temperatures.

\section {Phase fluctuations}

From action of (\ref{lag0}) we see that 
uniform shifts in $\theta$ do not cost any energy, and 
$\theta$ is the Goldstone mode of the superconducting state.

We now obtain the action for $\theta-$fluctuations 
coupled to fermions, setting $\eta_{\alpha} = \phi = 0$ (their saddle point 
values) in (\ref{lag0}) and (\ref{lag1}).
For slow spatial fluctuations, the deviation from the mean field
action, obtained from (\ref{lag0}) with $\theta = 0$, is given by 
\bea
\delta S[c^\dg,c,\theta] 
= \frac{1}{2 N} & & \int_0^{1/T} d\tau~\sum_{\vk,\vq,\sigma} 
c^{\dg}_{\vk,\sigma}(\tau) c_{\vk-\vq,\sigma}(\tau)
\left[i \partial_\tau -i (\xi_\vk - \xi_{\vk-\vq}) \right]
\theta_\vq(\tau) \\ \nonumber
 &-& \frac{1}{8 N} 
\int_0^{1/T} d\tau~\sum_{\vk,\vq,\vq',\sigma} c^{\dg}_{\vk,\sigma}(\tau) 
 c_{\vk-\vq-\vq',\sigma}(\tau) \theta_\vq(\tau) \theta_{\vq'}(\tau)
\left( \xi_\vk + \xi_{\vk-\vq-\vq'} - \xi_{\vk-\vq}-\xi_{\vk-\vq'} \right)
\label{Stheta}
\eea
Using $\theta_\vq(\tau+1/T)=\theta_\vq(\tau)$, an assumption
discussed in detail in Appendix B, and making a small $\vq$ expansion,
we arrive at
\be
\delta S[c^\dg,c,\theta] 
= \frac{1}{2 T N} \sum_{k,q,\sigma} c^{\dg}_{k,\sigma}
c_{k-q,\sigma} \theta_q
\left[ \omega_n -i \vv_{\vk\alpha} \vq_\alpha \right]
- \frac{1}{8 T N} \sum_{k,q,q',\sigma} c^{\dg}_{k,\sigma}
c_{k-q-q',\sigma} m^{-1}_{\alpha\beta}(\vk)
\vq_\alpha \vq'_\beta \theta_q \theta_{q'} 
\label{SthetaQ}
\ee
where $k \equiv (\vk, i \nu_m)$ and $q \equiv (\vq, i\omega_n)$ with
$\nu_m=(2 m+1)\pi T$ and $\omega_n= 2n\pi T$. We have used 
$\xi_{\vk-\vq}= \xi_\vk - \vv_{\vk\alpha}
\vq_\alpha + \frac{1}{2} m^{-1}_{\alpha\beta}(\vk) \vq_\alpha \vq_\beta +
\ldots$ where $\vv_{\vk\alpha} \equiv \partial \xi_\vk / \partial
\vk_\alpha$ is the velocity and $m^{-1}_{\alpha\beta}(\vk) \equiv 
\partial^2\xi_\vk/
\partial\vk_\alpha\partial\vk_\beta$ is the inverse mass tensor. 

\subsection{Neutral Systems}

For neutral systems we integrate out the fermions in (\ref{SthetaQ}),
using a cumulant expansion \cite{cumulant} controlled by small spatial and 
temporal gradients in $\theta$, leading to 
\be
S_{neutral}\left[\theta\right] = 
\frac{1}{T}\sum_{\vq,\omega_n} \frac{1}{8}\left[
- {\chi_0} \omega^2_n + {\Lambda_0}^{\alpha\beta}
\vq^\alpha \vq^\beta \right] \theta(\vq,i\omega_n) \theta(-\vq,-i\omega_n).
\label{neutralaction}
\ee
Here, ${\chi_0} \equiv - {1 \over T}
\la \rho(\vq,i\omega_n) \rho(-\vq,-i\omega_n) \ra$ 
is the mean field density-density correlation function given by
\bea
{\chi_0}(\vq,i\omega_n) &=& \frac{2}{\Omega} \sum_\vk (1-f-f') (u v' + v u')
\left[ {{u v'}\over{i\omega_n-E-E'}}-{{u' v}\over{i\omega_n+E+E'}}\right]
\nonumber \\
&+& \frac{2}{\Omega} \sum_\vk(f-f')(v v' - u u')
\left[ {{u u'}\over{i\omega_n-E+E'}}+{{v v'}\over{i\omega_n+E-E'}}\right]
\label{chi0}
\eea
and ${\Lambda_0}^{\alpha\beta}\equiv 
\frac{1}{\Omega}\sum_\vk m^{-1}_{\alpha\beta}(\vk)\la\hat{n}_\vk\ra
-{1 \over T} \la j^\alpha(\vq,i\omega_n) j^\beta(-\vq,-i\omega_n)\ra_0$ is 
the mean field phase stiffness, with
$\la \hat{n}_\vk \ra = \left( 1-\xi_\vk/E_\vk {\rm tanh}(E_\vk/2T) \right)$
and the paramagnetic current correlator 
\bea
{1 \over T} \la j_\alpha j^\ast_\beta \ra_0 &=&
\frac{2}{\Omega}\sum_\vk \vv_{\vk\alpha} \vv_{\vk\beta} (1-f-f') (v u'-u v')
\left[ {{u v'}\over{i\omega_n-E-E'}}+{{u' v}\over{i\omega_n+E+E'}} \right]
\nonumber \\
&+& \frac{2}{\Omega}\sum_\vk \vv_{\vk\alpha} \vv_{\vk\beta} (f'-f)(v v' + u u')
\left[ {{v v'}\over{i\omega_n+E-E'}}-{{u u'}\over{i\omega_n-E+E'}}\right]
\label{ds0}
\eea
$E$, $u$ and $v$ refer to standard BCS notation, $f = f(E)$ is the Fermi 
function, $E' \equiv E_{\vk-\vq}$ with $E \equiv E_{\vk}$
and similarly for other primed variables.
Analytically continuing $i\omega_n\to \omega+i\eta$,
and working at $T=0$ \cite{landausing}, we
obtain in the limit $\vq\to 0, \omega\to 0$,
the mean field superfluid stiffness
\be
{\Lambda_0}^{\alpha\beta}(T=0) 
= \frac{1}{\Omega} \sum_\vk m^{-1}_{\alpha\beta}\left(1-{{\xi_\vk}
\over{E_\vk}}\right) \equiv D^0_s(T=0) \delta_{\alpha\beta}
\label{bareds}
\ee
and the mean field compressibility
\be
{\chi_0}(T=0) = - \frac{1}{\Omega} 
\sum_\vk \frac{\Delta_\vk^2}{E^3_\vk} \equiv - \kappa
\label{mfchi}
\ee

We note that the effective action (\ref{neutralaction})
is appropriate for phase distortions whose energy
is smaller than the condensation energy 
$E_{\rm cond} = \frac{1}{8} D^0_s (\pi/\xi_0)^2 $.  
If this energy (density) is exceeded 
the $d$-wave BCS saddle point would become unstable.
This leads to the following restrictions in (\ref{neutralaction}):
$|\vq| < q_c \equiv \pi/\xi_0$ and $\omega_n < \omega_c$ 
with $\kappa \omega_c^2/8 = E_{\rm cond}$.
In the BCS limit,$E_{\rm cond} \simeq \Delta^2_d/v_{_F} k_{_F}$ which 
translates into $v_{_F} q_c \sim \omega_c \sim \Delta_d$.

We emphasize that
the coefficients in the phase action are {\it not}
the physical correlation functions. 
The gauge invariant correlation functions, obtained by including 
the effect of Gaussian phase fluctuations, are given by 
\be
\Lambda^{\alpha\beta}= {\Lambda_0}^{\alpha\beta}
+ {{{\Lambda_0}^{\alpha\mu}{\Lambda_0}^{\nu\beta}\vq^\mu \vq^\nu}
\over {\left[\omega^2_n{\chi_0} -
{\Lambda_0}^{\mu\nu} \vq^\mu \vq^\nu\right]}}.
\label{phyP0}
\ee
and
\be
\chi(\vq,i\omega_n)=
{{\vq^\alpha \vq^\beta {\Lambda_0}^{\alpha\beta} {\chi_0}}
\over{\vq^\alpha\vq^\beta {\Lambda_0}^{\alpha\beta}
- \omega^2_n {\chi_0}}}.
\label{phyChi0}
\ee
as shown in Appendix C. 

From (\ref{phyP0}) and (\ref{phyChi0}) we see that {\em Gaussian} phase 
fluctuations do not affect transverse correlation functions. In
particular, the superfluid stiffness is unrenormalized. 
However, longitudinal correlations {\em are} affected in general. 
While $\Lambda_0$ does not satisfy
the $f-$sum rule, $\Lambda$ does, 
which implies restoration of gauge invariance.
Further, from (\ref{phyChi0}) we can see that
$\chi$ has a pole for $\vq \to 0$ unlike ${\chi_0}$, which leads to 
a collective mode which we will discuss in the next Section. 
However, the static compressibility given by 
$-\chi(\vq\to 0,\omega_n=0)$ is unaffected at the Gaussian level. 

We clearly see that the gauge-invariant 
$(\vq,\omega)$ dependent correlation functions are different from
the mean field correlations which appear as coefficients of the phase action.
This is not surprising since the phase variable in fact
contributes to the physical longitudinal correlation functions and modifies
the mean field result. This is true even for charged systems as will be shown 
below. 

\subsection{Charged systems}

In charged systems we have to take into account the
long range Coulomb interaction with energy density
\be
\frac{1}{\Omega} H_{\rm coulomb}= 
\frac{1}{2 N} \sum_\vq V_\vq \rho_\vq \rho_{-\vq}
\label{Vcoulomb}
\ee
where $\rho_\vq \equiv \frac{1}{\Omega} 
\sum_{\vk,\sigma} c^{\dg}_{\vk+\vq,\sigma} c_{\vk,\sigma}$  is the
electron density. In anisotropic layered systems $V_\vq$
is given by \cite{fetter74,fertig91}
\be
V_\vq = \frac{2 \pi e^2 d_c}{q_\pll \epsilon_b}
\left[\frac{\sinh(q_\pll d_c)}{{\rm cosh}(q_\pll d_c)-\cos(q_\perp d_c)}\right]
\label{coulomb}
\ee
where $q_\pll,q_\perp$ denote in-plane and $c-$axis components of $\vq$,
$d_c$ denotes the mean interlayer spacing, and 
$\epsilon_b$ is the background dielectric constant.
We assume $q_\pll a \ll 1$ always, where the in-plane lattice spacing $a=1$.
For $q_\pll d_c, q_\perp d_c \ll 1$ this reduces to the ordinary 3D result:
$V_\vq= 4 \pi e^2/(\epsilon_b q^2)$. 

We take our Hamiltonian to be $K + H_{\rm pair} + H_{\rm coulomb}$.
Since the short range attraction of $H_{\rm pair}$
is important for small center-of-mass momentum while
the Coulomb effects are important for 
small momentum transfer, we believe the breakup of
the actual interaction in this manner is physically sensible and
does not lead to any ``overcounting''.

The Coulomb interaction can now be decoupled
using a field $\cV_\vq(\tau)$ as
\be
\exp\left[-\sum_{\vq>0} V_\vq \rho_\vq(\tau)
\rho_{-\vq}(\tau)\right]
= \int \cD \left( \cV_\vq (\tau) , \cV^*_\vq (\tau) \right)
\exp\left[- \sum_{\vq>0} \frac{1}{V_\vq} \cV^*_\vq \cV_\vq + 
i \sum_{\vq>0} 
\left(\cV^*_\vq \rho_{-\vq} + \cV_\vq \rho_{\vq} \right)\right]
\label{decoup_Cou}
\ee
To integrate over independent modes, since $\cV_{-\vq}= \cV^{\ast}_\vq$,
we only sum over $\vq>0$. 
The last term in (\ref{decoup_Cou}) can be recast as 
$\sum_{\vr} \cV(\vr) \rho(\vr)$.
Thus the density $\rho$ couples to the scalar field $\cV$ in the same
way as it couples to $\partial_\tau\theta$ in (\ref{Stheta}). 

On integrating out the fermions 
we arrive at an effective action for the 
phase $\theta_q$ and the scalar potential $\cV_q$, 
given by
\be
S[\theta, \cV] = \frac{1}{8 T} \sum_{\vq,i\omega_n} \left[ \begin{array}{ll}
\theta^*(\vq,i\omega_n) & \cV^*(\vq,i\omega_n) \end{array} \right] 
~{\cal M} ^{-1}~
\left[ \begin{array}{l} \theta(\vq,i\omega_n) \\ \cV(\vq,i\omega_n)\end{array} 
\right] \label{VTheta}
\ee
with
\be
{\cal M}^{-1} = 
\left( \begin{array}{ccc}
-\omega^2_n {\chi_0} + {\Lambda_0}^{\alpha\beta}\vq^\alpha \vq^\beta & & 2 i
\omega_n {\chi_0} \\ & & \\ 
- 2 i \omega_n {\chi_0}^* & & 4 (-{\chi_0} + V_\vq^{-1}) \end{array} \right) 
\label{matrix}
\ee
where ${\chi_0}^*=\chi_0(-\vq,-i\omega_n)$. 
Integrating out the field ${\cal U}$ leads to the action
\be
S_{charged}[\theta] = \frac{1}{8 T}\sum_{\vq,\omega_n} \left( -\omega_n^2
\chi_0^{\small RPA}
+ {\Lambda_0}^{\alpha\beta}
\vq^\alpha\vq^\beta \right)
\theta(\vq,i\omega_n) \theta(-\vq,-i\omega_n)
\label{chargedaction}
\ee
where the mean field charged system density correlator 
$\chi_0^{\small RPA}=\chi_0/(1-V_\vq\chi_0)$. 
This form of the phase
action is independent of the order parameter symmetry and has been obtained 
earlier for $s$-wave SCs \cite{tvr89}. We note that this
form differs considerably from that assumed in Ref.~\cite{emery95}, 
where the {\em physical} $(\vq,\omega)$-dependent longitudinal dielectric 
function appears as a coefficient in the action.  As emphasized earlier,
{\em physical} longitudinal correlations cannot appear as coefficients of 
the phase action.

The regime of validity for
the above action is $\vq \lesssim\pi/\xi_0$ as for neutral
systems (see discussion below (\ref{mfchi})). 
The frequency cutoff is given by
$|\chi_0^{\small RPA}(q) \omega^2_n| \lesssim E_{\rm cond}$ 
and thus depends on $\vq$.
In particular, for $\vq\to 0$, the action remains valid even at 
frequencies larger than the gap $\Delta_d$ 
and can be used to obtain the $\vq \to 0$ plasma mode.

The gauge-invariant correlations for the charged system are obtained from
the neutral system results (\ref{phyP0}) and (\ref{phyChi0}) by replacing 
${\chi_0} \to \chi_0^{\small RPA}$, as discussed in Appendix C. 

\section{Plasmons}

In the previous Section we have found phase actions of the form
$S[\theta]=\frac{1}{T}\sum_{\vq,\omega_n}
M^{-1}_\theta |\theta(\vq,\omega_n)|^2$ for
neutral systems (see eq.~(\ref{neutralaction})) and for charged
SCs (see eq.~(\ref{chargedaction})).
The dispersion of the collective phase mode is defined by
${\cal R}e M^{-1}_\theta (\vq,\omega)=0$. 
For neutral systems, we note that this condition is
identical to demanding a pole in physical density-density correlation 
$\chi$ given in (\ref{phyChi0}). 
Phase and density fluctuations are thus coupled and share the pole of 
the collective mode. 
This is true even for charged systems, 
where the plasma frequency, $\omega_p$,
corresponds to the pole of the physical density correlator and is
given by $\lim_{\vq\to 0} {\cal R}e \chi^{-1}(\vq,\omega_p)=0$.

The phase action is valid for all frequencies such that
$\omega^2 \lesssim E_{\rm cond} V_{\vq}$
and thus is valid even for large
frequencies for $\vq\to 0$.
If the plasmon is at finite frequency for $\vq\to 0$,
Landau singularities do not occur at finite temperatures.
One can thus use this action to obtain the
$\vq\to 0$ plasma mode at zero and finite temperatures.
Further, to have a sharp plasmon the damping must be relatively
small: ${\cal I}m M_\theta \ll \omega_p$.

In this section, we first briefly consider neutral systems followed 
by a discussion of charged systems. For charged systems, we study
the in-plane plasma mode and then consider $c-$axis plasmons for systems 
with a finite $c-$axis superfluid stiffness. Our discussion of the
c-axis plasmon is to a large extent independent of the details of any 
$c-$axis model. We also make an estimate of 
the $c-$axis plasma frequency for Bi2212 obtained within this phase 
action and compare it with experiment.

\subsection{Neutral Systems}

For neutral systems we have $M_\theta^{-1} = 
\frac{1}{8}(\Lambda_0^{\alpha\beta}
\vq_\alpha\vq_\beta-\chi_0 \omega^2_n)$. At $T=0$, continuing to real
frequency, the collective mode frequency obtained from above
is given by $\omega(q)=cq$ where the sound
velocity $c\equiv \sqrt{D^0_s/\kappa}$. This reduces to the
standard result $c=v_{_F}/\sqrt{d}$
in the weak coupling limit in the continuum,
where $d$ is the spatial dimension and $v_{_F}$ is the Fermi velocity. 
At finite temperatures, there are Landau singularities in $D^0_s$ and
$\chi_0$ as seen from (\ref{chi0}) and (\ref{ds0}), which prevents
us from taking the $\vq \to 0, \omega \to 0$ limit.

\subsection{Charged Systems}

In the action (\ref{chargedaction}), we can set $\chi_0^{\small RPA}
\to -1/V_\vq$
in the limit $\vq \to 0$.
Analytically continuing to real frequency, and setting ${\cal R}e
M^{-1}_\theta(\vq,\omega_p)=0$, we obtain
\be
\lim_{\vq\to 0} \left[
q^\alpha q^\beta {\cal R}e \Lambda_0^{\alpha\beta} (\vq,\omega_p(\hat{q}))
-\frac{1}{V_\vq}\omega_p^2(\hat{q}) \right]= 0,
\ee
where $\omega_p$ depends on the direction of propagation
(in-plane or $c-$axis) in anisotropic systems. 

We first consider the in-plane plasmon. For 
$\vq\to 0$ ($\hat{\vq}$ in-plane) and finite $\omega$ we can write
$\Lambda^{\alpha\beta}_0 = \delta_{\alpha\beta}~(-i
\epsilon_b~\omega \sigma(\omega)/e^2)$ where $\sigma\equiv
\sigma^\prime+i\sigma^\dprime$ is the in-plane optical conductivity 
(including the effects of Gaussian phase fluctuations). The background
dielectric constant enters in this definition of $\sigma$ since the 
conduction electrons are affected by the electric field which has been
screened by the background.
We thus see 
that 
\be
\lim_{\vq\to 0} V_\vq \vq^2 \frac{ \epsilon_b
\sigma^\dprime_{ab}(\omega_{p})}{e^2} = 
4\pi \sigma^\dprime_{ab}(\omega_{p}) = 
\omega_{p} 
\label{plasma}
\ee
$\sigma(\omega)$ thus governs the location and damping of the plasma mode
through a self-consistent equation. This equation is identical to
demanding a zero in the real part of the longitudinal dielectric 
constant ($\epsilon$), since $\epsilon=\epsilon_b+4\pi i \sigma/\omega$ in a 
{\em gauge invariant theory}. 

At this stage it is instructive to compare (a) the physical plasma
frequency $\omega_p$, (b) the conductivity sum rule plasma 
frequency $\omega_p^*$ defined by
\be
\int_{0}^{\infty} d\omega~\sigma^\prime
(\omega)= \frac{e^2}{\epsilon_b} \frac{\pi}{2\Omega}
\sum_\vk m^{-1}(\vk) \la n(\vk)\ra \equiv \frac{\omega_{p\alpha}
^{* 2}} {8},
\label{sumrule}
\ee
and (c) the superfluid plasma frequency $\omega_{ps}$ 
defined by
$\omega^2_{ps}(T)\equiv 4\pi e^2 D_{s}(T)/\epsilon_b$, where $D_{s}(T)$ is the
$T$-dependent superfluid stiffness related to $\lambda(T)$, 
the penetration depth \cite{lambda}.
With this definition, the 
real part of $\sigma$ can be represented as
$\sigma^\prime(\omega,T)=(\omega_{ps}^2(T)/4)\delta(\omega)
+\sigma_{reg}(\omega,T)$ 
where $\sigma_{reg}$ is the regular part.
Finally, we have the Kramers-Kr\"onig relation for $\sigma(\omega)$:
\be
\sigma^\dprime(\omega)= \frac{1}{\pi}{\cal P}\int_{0}^{\infty}
d\omega'~\sigma^\prime(\omega') \frac{2\omega}{\omega^2-\omega'^2}
\label{KK} 
\ee
We will now try to use these relations and the structure of 
$\sigma^\prime(\omega)$ to obtain direct information about the behavior of 
the plasma mode, which is not directly seen in an optical conductivity 
measurement. 

Conventional {\em clean} 3D $s$-wave SCs at $T=0$ have a very large superfluid 
stiffness which can be inferred from penetration depth measurements, and
little spectral weight at higher energies. 
Ignoring interband transitions, we can then take $\sigma^\prime(\omega)\approx
(\omega_{ps}^2(0)/4)\delta(\omega)$ which leads 
to $\omega_p^*=\omega_{ps}(0)$ from (\ref{sumrule}). 
It also implies $\sigma^\dprime(\omega)= \omega_{ps}^2(0)/4\pi\omega$ 
through the Kramers-Kr\"onig relation (\ref{KK}). 
Using (\ref{plasma}) we then get
$\omega_p=\omega_{ps}(0)$, a large
plasma frequency. In the presence of weak disorder and at finite $T$,
$\omega_{ps}$ decreases and the spectral weight in $\sigma^\prime(\omega)$ 
redistributes, leading to finite $\sigma(\omega)$ over energy scales 
$\tau_{\rm tr}^{-1},\Delta$ which are the quasiparticle transport lifetime 
and SC gap respectively. Since $\omega_p \gg \tau_{\rm tr}^{-1},\Delta$ 
to begin with, it is unaffected by this low energy redistribution. This 
is easy to
see from (\ref{KK}) above for $\sigma^\dprime(\omega)$ where we can set
$\omega{'} \approx 0$ in the denominator for the region of interest, and
this along with (\ref{sumrule}) and (\ref{plasma}) leads to $\omega_p=
\omega^*_p$, which is insensitive to the spectral weight 
redistribution. 
Further, the small $\sigma^\prime(\omega_p)$ implies a sharp plasmon in this
case. Thus for conventional $s$-wave SCs, we finally arrive at $\omega_p 
\approx \omega_p^*\geq
\omega_{ps}(T)$. The last 
relation is satisfied as an equality only in a Galilean invariant system 
at $T=0$. 

For the cuprate superconductors, the in-plane $\sigma^\prime(\omega,T=0)$ 
has the following features: $(i)$ a condensate contribution 
$(e^2\pi D^0_s(T=0)/\epsilon_b)\delta(\omega)$ and 
$(ii)$ absorption by quasiparticles 
which has low frequency spectral weight (in a $d$-wave SC) with features
around twice the maximum gap followed by other higher energy features.
The condensate contribution {\em along} with the large low energy spectral 
weight coming from quasiparticles is expected to lead to the large plasma 
frequency, as in the case of conventional SCs above. Ignoring interband 
transitions in calculating $\omega^*_p$, we then arrive at $\omega_p \approx 
\omega^*_p >\omega_{ps}$.

The {\em normal state} in-plane plasma frequency has been 
measured to be large ($\sim 1 eV$) in the cuprates \cite{uchida91} while 
the spectral weight rearrangement in $\sigma^\prime$ in going from the 
normal state to the SC 
state is over smaller energy scales \cite{orenstein90}. The high energy 
normal state plasmon is thus expected to smoothly go over into a high energy 
SC state plasmon expected from our above discussion, similar to
conventional SCs.

\subsection{Josephson Plasmons along c-axis}

To study the $c-$axis plasmon we assume a non-vanishing $c-$axis 
stiffness $D_{_\perp}$.
We set $\hat{\vq}$ along the $c-$axis in (\ref{plasma}),
leading to $\omega_{p,c} = 4\pi \sigma_c^\dprime(\omega_{p,c})$.
We know $\omega_{ps,c}$ is small in the high $T_c$ systems 
as seen from the large $c$-axis penetration depths, and $\sigma_c^\prime
(\omega)$ 
is measured to be very small over a very large energy range \cite{uchida92}.
This is partly due to the form of the $c$-axis dispersion which is 
proportional to $(\cos k_x-\cos k_y)^2$\cite{okandersen}, 
so that nodal quasiparticles which lead to low energy spectral weight 
in-plane have a much smaller contribution along the $c$-axis. 
Thus at $T=0$, the only ``free carriers'' come from the condensate, and 
$\omega_{p,c}^{*} \simeq \omega_{ps,c}$. On using
(\ref{KK}) and (\ref{sumrule}), this leads to $\omega_{p,c}(T=0)=\omega_{ps,c}
(T=0)$. With increasing temperature, the spectral weight in
$\sigma_c^\dprime$
gets transferred to very high energies \cite{basov99}. The plasma 
frequency then continues to be given by
$\omega_{p,c}(T)=\omega_{ps,c}(T)$, 
which decreases with increasing $T$ and vanishes above $T_c$ as seen in 
experiment \cite{uchida92,homes93,basov94}. Thus the 
$c$-axis plasma oscillations are seen only below $T_c$,
justifying the use of the term ``Josephson plasmon''.

A model which assumes disordered quasiparticle transport along the 
$c$-axis \cite{dassarma98-1} and a model which only permits pair tunneling 
\cite{ioffe99} both appear to lead to this behavior for $\omega_{pc}(T)$. 
Presumably
the main effect of disorder in the first model is to suppress single
particle tunneling leading to pair tunneling as the dominant process
below $T_c$; it then appears that the two models are 
similar in spirit. The disorder scale required to reproduce the 
experimental results in the first model appears to be large, of the scale 
of the one particle tunneling bandwidth.
One reason for the large disorder scale could be that the
$c$-axis dispersion in this calculation has been chosen to be independent 
of $k_x,k_y$, while the actual $(\cos k_x-\cos k_y)^2$ dependence would 
already suppress single-particle tunneling near the nodes in the clean case. 
This might then lead to a smaller disorder scale required to suppress this 
tunneling process completely.

We finally proceed to compare the $c$-axis plasma frequency with the 
stiffness obtained from penetration depth experiments in Bi2212. We
consider only the in-phase $c$-axis plasmon and ignore the ``optical mode''
corresponding to out-of-phase fluctuations arising from the bilayer
structure, which is expected to be at a higher energy 
\cite{griffin95,dassarma98-2}. 
This leads to 
\be
\omega^2_{p,c} = \frac{4\pi e^2}{\epsilon_b} D_{_\perp}=\frac{c^2}
{\epsilon_b\lambda^2_c}
\ee
where $c$ is the velocity of light and $\lambda_c$ is the low temperature 
$c-$axis penetration depth. $\lambda_c$ in Bi2212 has been 
measured \cite{cooper90} to be about $100$ microns. Using this
and setting $\epsilon_b \approx 10$, we get $\omega_{p,c} 
\approx 7 K$. 
This is in reasonable agreement with $\omega_{p,c} \approx 
8$-$10 K$ extracted from 
experiment\cite{tsui96,mallozzi97,kadowaki98} given experimental
errors, and uncertainties in the estimate of $\epsilon_b$.

\subsection{Plasmon dispersion}

In order to understand the plasmon dispersion and the variation of the 
plasma energy with direction of propagation, we consider a simplified 
model for the in-plane and out-of-plane conductivity. 
Since $\omega_{p, ab} \approx \omega^*_p$ and is
large for the in-plane plasmon, we set the in-plane conductivity 
$\sigma^\prime(\omega)=(\omega^{2 *}_p/4) \delta(\omega)$, consistent with 
the conductivity
sum rule, which then reproduces the large in-plane plasma frequency. 
We emphasize that this simplified form for the in-plane conductivity 
is valid {\em only} at large frequency, for 
studying the high energy in-plane plasmon.
It is {\em not valid} for 
considering low energy in-plane physical observables, such as the
superfluid stiffness.
Using this form for the real conductivity, 
$\sigma_{ab}^\dprime(\omega)=\omega_p^{* 2}/4\pi\omega$.
The $c$-axis conductivity is given by $\sigma^\prime_c=
(\omega^2_{ps,c}/4)\delta(\omega)$ as discussed in the last Section, 
which leads to $\sigma^\dprime_c(\omega)=\omega_{ps,c}^{2}/4\pi\omega$.
This simplified model is tailored to capture the correct
plasma frequency for propagation along the in-plane and $c$-axis
directions. For an arbitrary direction of propagation, we expect to
obtain a reasonable interpolation. 

\begin{center}
\begin{figure}
\leavevmode
\epsfig{figure=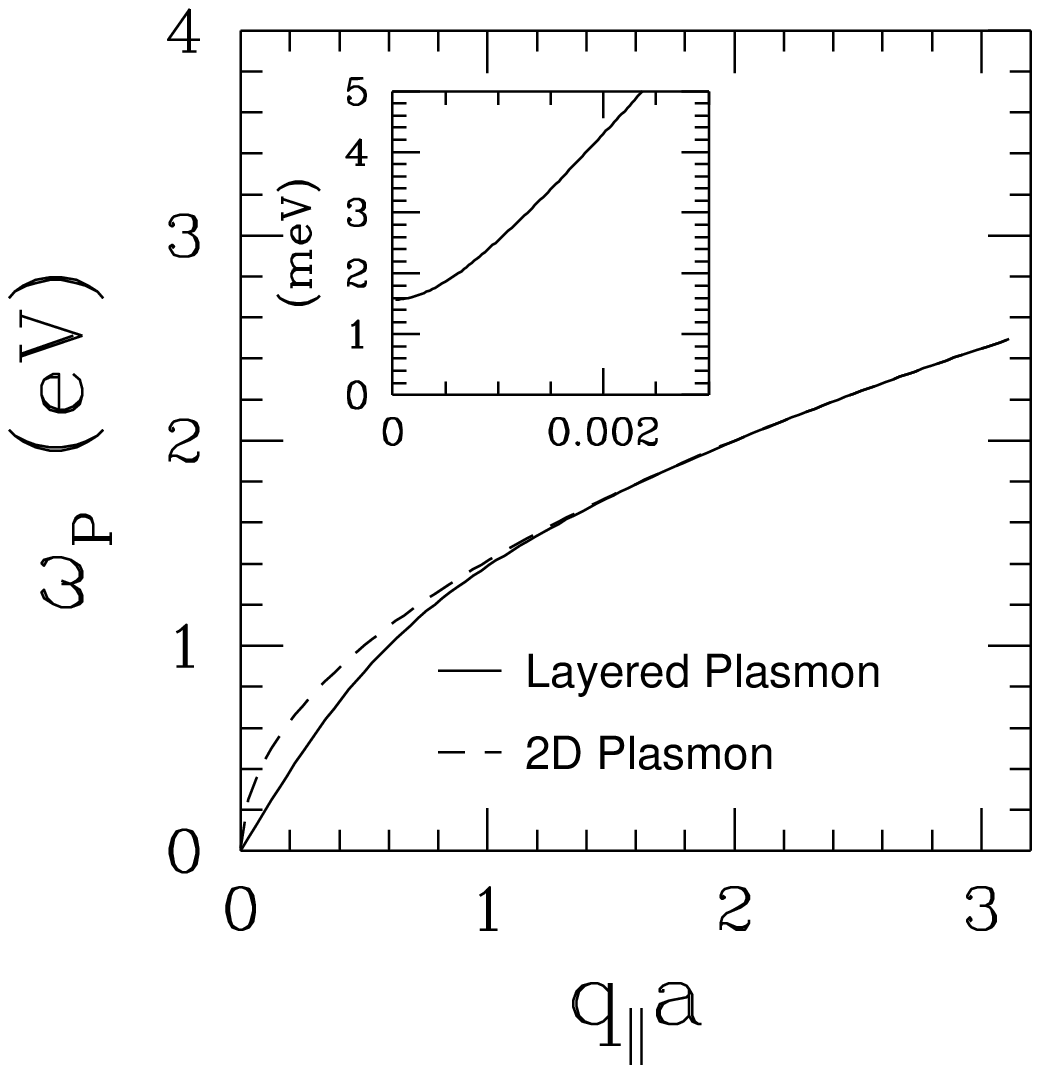,width=14cm,angle=0}
\caption{The plasmon dispersion in a layered system (Bi2212), 
as a function of $\vq_{_\pll}$ with $q_{_\perp}=\pi/d_c$ (see text for 
details and parameters used). 
The $\sqrt{q_{_\pll}}$ dispersion in the 2D limit is plotted for comparison. 
The dispersion appears nearly acoustic for small $q_{_\pll}$ due to 
the {\em very small} Josephson plasma frequency ($\sim 10 K$) but rapidly 
crosses over to high energies ($\sim eV$) with increasing $q_{_\pll}$.
Inset shows the behavior of the dispersion (in $meV$) for $q_{_\pll}\to 0$.
}
\label{abplasmon}
\end{figure}
\end{center}

For general $\vq$, in the absence of detailed information on the 
$\sigma(\vq,\omega)$,
we assume that the conductivity is independent of $\vq$. This appears
to be a reasonable assumption at high energy, and leads to 
the plasma frequency being given by
\be
\omega_p(\vq) = 
{V_\vq \epsilon_b \over e^2}~~\left(\sigma_{ab}^{\dprime}(\omega_p)
q_{_\pll}^2+ \sigma^\dprime_c(\omega_p) q_{_\perp}^2 \right)
\ee
The plasma frequency for $|\vq| \rightarrow 0$ is a function of the angle
of propagation, $\psi$, measured with respect to the $ab$-plane and given 
by $\omega_p^2(\psi)= \left(\omega_p^{* 2} \cos^2(\psi)+\omega_{ps,c}^2
\sin^2(\psi)\right)$ and varies smoothly from the low energy Josephson
plasmon for $c$-axis propagation to the high energy plasmon in-plane.
In order to examine the plasmon dispersion
in a particular case, we consider the limit $q_{_\perp}=\pi/d_c$ and
study $\omega_p$ as a function of $(\vq_{_\pll} a)$.
Normal state data for LSCO \cite{uchida91} shows
$\omega_{p,\pll}(q_\perp=0,\vq_{_\pll}\to 0) \sim 1 eV$; we assume
a similar value for Bi2212 with 
$\omega_{p,\pll}(q_\perp=0,\vq_{_\pll}\to 
0)\approx\omega_p^{*}$. Using $d_c/a \approx 4$
and $\omega_{pc}=\omega_{ps,c}\sim 1.0 meV$
\cite{tsui96,mallozzi97,kadowaki98}, we plot the plasma frequency in
Fig.\ref{abplasmon}. This crosses over from a low energy 
Josephson plasmon 
for $q_{_\pll}\to 0$, corresponding to $c$-axis propagation, to a nearly two 
dimensional plasmon dispersing as $\sqrt{q_{_\pll}}$ at larger $q_{_\pll}$.  
At small $q_{_\pll}$, the dispersion is known to be acoustic for 
$\omega_{ps,c}=0$ \cite{fertig91}.
It appears to be acoustic in Bi2212 at small $q_{_\pll}$ (see Fig.
\ref{abplasmon}) due the 
extremely small value of $\omega_{ps,c}$, but finally levels off leading
to a finite Josephson plasmon gap for $q_{_\pll}\to 0$ as shown in the 
inset in Fig.\ref{abplasmon}.
The 2D $\sqrt{q_{_\pll}}$ dispersion is obtained
mathematically in the limit of large $d_c/a$ and is given by
$\omega_{p,2D}(\vq_\pll)=(\omega^{*}_p/\sqrt{2})\sqrt{q_{_\pll}d_c}$
and we plot this in Fig.\ref{abplasmon} for comparison.
It must be emphasized that plasmon damping would be important in the 
real system and would have non-trivial dependence on the angle of
propagation. The sharp plasmon we obtain in the above cases is an 
artifact of our simplified model for the conductivity.

\section{Quantum XY Model} 

The superfluid stiffness obtained above
$ D^0_{s}(T) = \frac{1}{\Omega}\left[ 
\sum_\vk m^{-1}_{xx}\left( 1-\xi_\vk/E_\vk\right) 
- 2\sum_\vk \vv_x^2 \left(-\partial f/\partial E_\vk \right)\right]$
is unaffected by Gaussian phase fluctuations. 
Corrections to this result are unimportant in conventional
superconductors which have a large coherence length and 
a large $D^0_s(T=0)$.
However, as we show below, effects beyond the Gaussian approximation
could lead to large corrections in systems with a small coherence
length and small $D^0_s$. To study such effects we derive in this Section
a quantum XY model and analyze quantum and thermal phase fluctuations 
in the following Section. For clarity of presentation, we outline the 
derivation for a $d$-dimensional isotropic system; the generalization to the
anisotropic case is straightforward.

The quantum XY model describes the dynamics of the phase variables
$\theta_\vR(\tau)$ defined
on a lattice with lattice spacing $\xi_0$, the coherence length.
The simplest action periodic under  
$\theta_\vR(\tau)\to \theta_\vR(\tau)+2\pi$ is given by 
\be
S_{XY}[\theta] = \sum_{\vQ,\omega_n} A(\vQ) 
\omega^2_n |\theta(\vQ,i\omega_n)|^2 +
B \int_0^{1/T}d\tau~\sum_{\la \vR,\vR' \ra}  
\left[1- \cos (\theta_\vR(\tau) - \theta_{\vR'}(\tau))\right],
\label{xy1}
\ee
where $\la \vR,\vR' \ra$ are neighboring sites.
It is important to emphasize that, given 
the $\cos (\theta_\vR - \theta_{\vR'})$ form,
there are no constraints on the spatial gradient of the phases 
defined on the coarse-grained scale of $\xi_0$. 
This is in contrast to the Gaussian action (\ref{chargedaction}), derived 
on the scale of the microscopic lattice spacing $a$,
which could only describe slow spatial fluctuations of the phase
whose energy did not exceed the condensation energy.

Our task now is to determine the coefficients $A(\vQ)$ and $B$ of
this effective action. Unlike in some other cases \cite{perturb}
it is not possible to directly derive the quantum XY action from
the underlying fermionic Hamiltonian, since the cumulant expansion
we used to derive effective phase actions was
controlled by the smallness of spatio-temporal gradients in $\theta$.
We therefore proceed as follows: we compare the action (\ref{xy1})
in the limit of slow spatial variations on the scale of the coherence 
length and match coefficients with those of the Gaussian action:
\be
S[\theta] = \frac{1}{8 T}{\sum_{\vq,\omega_n}}'' {{\omega^2_n a^d
}\over
{V(\vq)}}|\theta(\vq,i\omega_n)|^2 + \frac{1}{8}\int_0^{1/T}d\tau
\sum_{\vr,\alpha}
D^0_s a^{d-2} \left[\theta_\vr(\tau)-\theta_{\vr+\alpha}(\tau) \right]^2
\label{Gaussian}
\ee
where $V(\vq)$ is the generalized $d$-dimensional Coulomb interaction.
For the $(\vq,i\omega_n)$ of interest, we have set ${\chi_0^{\small RPA}} 
\approx -1/V(\vq)$ and $\Lambda_0 = D^0_s(T)$.
The double prime on the summation denotes momentum and 
frequency cutoffs
\be
|\vq| < q_c \equiv \pi/\xi_0 \ \ \ {\rm and} \ \ \ 
\omega_n^2 \leq (2\pi n_c T)^2 \equiv D^0_s \left(\frac{\pi}{\xi_0}\right)^2
V_\vq
\label{cutoffs}
\ee
which arise from demanding that the energy cost of the terms in 
(\ref{Gaussian}) to be less than the condensation energy $E_{\rm cond}
=\frac{1}{8} D^0_s (\pi/\xi_0)^2$. 
The $\vq$ cutoff can also be viewed as a representation of the
$\vq$-dependent stiffness being roughly constant ($\approx D^0_s$) for 
$\vq<q_c$ and decreasing to zero for $\vq>q_c$.

In arriving at (\ref{Gaussian}), we have Fourier transformed the gradient 
term from the $(\vq,i\omega_n)$ to $(\vr,\tau)$ variables. While the
above $\tau$-local form of this term is true when {\em all}
Matsubara frequencies are present, we now determine its regime of 
validity given the frequency cutoff in (\ref{cutoffs}). 
With the cutoff in (\ref{cutoffs}), the gradient term on Fourier 
transforming is given by
\be
\frac{T}{8} \sum_\vq D^0_s a^d \vq^2 \int_0^{1/T}~d\tau~d\tau' \theta(\vq,\tau)
\theta(\vq,\tau') K(T(\tau-\tau'))
\ee
In terms of the dimensionless quantity $z=\tau T$ the kernel is given by
$K(z) = \sin\left[ (2 n_c+1) \pi z \right]/\sin(\pi z)$
where $n_c \equiv n_c(\vq)$ given by (\ref{cutoffs}).
The kernel $K(z)$ is periodic in $z$, $K(z+1)=K(z)$, and it is sharply
peaked around $z=0$ for large $n_c$. The
width of the peak can be estimated from the first zero of $K(z)$ as
$z_0 = 1/(2n_c+1)$. For $n_c \stackrel{_>}{_\sim} 10$, $z_0 \ll 1$
which is true in the low temperature regime that we shall be interested 
in. We thus approximate $K(z)$ as a delta function in ``time'' and
work with a local-$\tau$ action in (\ref{Gaussian}).

To determine the coupling $B$ in (\ref{xy1}), we consider static
$\theta$ configurations and apply a small external twist $\Phi$ 
which will be distributed uniformly over the system. 
For the Gaussian model (\ref{Gaussian}), with lattice spacing $a=1$
and $N=L^d$ sites, the phase twist per link is $(\Phi/L)$, while
for the the XY model (\ref{xy1}), on the ``coarse grained'' lattice
with lattice spacing $\xi_0$ and $(L/\xi_0)^d$ sites, there is a  
larger phase gradient $\Phi/(L/\xi_0)$.
Since the total energy cost for this phase twist is the same
in the two cases, one obtains
$D^0_s a^{d-2}~\big(\Phi/L\big)^2~L^d/8 =  
B~\big(\Phi\xi_0/L a\big)^2~\big(L a/\xi_0\big)^d/2$, 
which leads to $B= D^0_s \xi_0^{d-2}/4$. 

Similarly, for a $\tau$-dependent phase fluctuation at a 
frequency $\omega_n$, $(L/\xi_0)^d$ phases contribute in the
``coarse grained'' XY model, as opposed to $L^d$ in the Gaussian case.
We thus get $A(\vQ)\omega^2_n(L a/\xi_0)^d = (\omega^2_n a^d/8V_\vq) 
L^d$ on equating the first term in the two actions. This leads to
$A(\vQ)=\xi_0^d/8V_\vq$. 
Finally, noting that the momentum $\vQ=\vq \xi_0/a$ since the
distances in the ``coarse
grained'' lattice are in units of $\xi_0$, this can also be written as
$A(\vQ) = \xi_0^d/8\tilde{V}(\vQ)$, where $\tilde{V}(\vQ) = 
V(\vQ~a/\xi_0)= V(\vq)$.

Having obtained $A(\vQ)$ and $B$ in the isotropic $d$-dimensional
case, we generalize to anisotropic systems. We work with a layered 3D
system with lattice spacing $a=1$ in-plane and $d_c$ along the
$c$-axis. The in-plane coherence length $\xi_0 \gg a$ and the
$c$-axis coherence length $\xi_{_\perp}=d_c$. In this case, since
$\xi_{_\perp}=d_c$, we proceed
exactly as above but coarse grain {\em only} the in-plane variables. 
Denoting the Gaussian model stiffness 
by $D^0_{_\pll}$ in-plane and $D^0_{_\perp}$ along the $c$-axis, we arrive 
at the final action
\bea
S[\theta] = \frac{1}{8 T}{\sum_{\vQ,\omega_n}}' \frac{\omega^2_n \xi_0^2
d_c}
{\tilde{V}_\vQ} \theta(\vQ,\omega_n) \theta(-\vQ,-\omega_n)
& + & \frac{D^0_{_\pll} d_c}{4} \int_0^{1/T}~d\tau \sum_{\vr,\alpha=x,y} 
\left(1-\cos[\theta(\vr,\tau)-\theta(\vr+\alpha,\tau)] \right) \\ \nonumber
& + & \frac{D^0_{_\perp} d_c}{4}\left(\frac{\xi_0}{a}\right)^2
\int_0^{1/T}~d\tau \sum_{\vr} 
\left(1-\cos[\theta(\vr,\tau)-\theta(\vr+\hz,\tau)] \right).
\label{xy2}
\eea
where $\tilde{V}(\vQ)=V(\vQ~a/\xi_0)$ 
and $V(\vQ)$ is the Coulomb interaction 
for layered systems given in (\ref{coulomb}).
While all $|\vQ| \leq \pi$ contribute in (\ref{xy2}) above, the prime
on the summation denotes the Matsubara cutoff consistent with 
(\ref{cutoffs}). The couplings in this action depend crucially on $\xi_0$ 
and we shall examine the consequences of this below.

\section{Renormalization of the stiffness}

The quantum XY model action has both longitudinal fluctuations and
transverse (vortex) excitations. Near a finite temperature phase
transition, the dynamics is unimportant and we recover the classical 
XY model with the possibility of a phase transition in the 3D-XY
universality class. In this Section, we examine low temperature properties
and the effect of quantum dynamics. We ignore vortex-antivortex pair 
excitations since these have a core energy cost and would be exponentially 
suppressed at low temperatures. We deal with the longitudinal fluctuations 
within a self consistent harmonic approximation (SCHA).

To examine the low temperature in-plane properties, we assume 
$D^0_{_\perp}=0$ in
(\ref{xy2}) since it is very small in highly anisotropic systems with 
a large $\lambda_{_\perp}$.
The $c$-axis stiffness would become important if $\lambda_{_\perp} 
\lesssim \lambda_{_\pll}(\xi_0/a)$, which implies $D_{_\pll}
\lesssim D_{_\perp}(\xi_0/a)^2 $, leading to the $c$-axis contribution
being important in (\ref{xy2}).
For Bi2212, detailed calculations, which we omit here, show that it does 
not affect our in-plane results. 

The SCHA \cite{schakra88,roddick95} is carried out by replacing the above 
action by a trial harmonic theory with the
renormalized stiffness $D_{_\pll}$ chosen to minimize the free energy of the
trial action. This leads to $D_{_\pll} = D^0_{_\pll} 
\exp(- \la\delta\theta^2\ra/2)$ 
where $\delta\theta \equiv (\theta_{\vr,\tau}-\theta_{\vr+\alpha,\tau})$
and the expectation value is evaluated in the renormalized harmonic theory. 
Explicitly, we get
\be
\la\delta\theta^2\ra = 2 T \int_{-\pi}^{\pi}~\frac{d^3\vQ}{(2\pi)^3} 
\sum^{n_c}_{n=-n_c} \frac{\epsilon_\vQ}{\omega^2_n 
\xi_0^2 d_c/\tilde{V}_\vQ + D_{_\pll} d_c \epsilon_\vQ}
\label{rmstheta}
\ee
where $\epsilon(\vQ)=4-2\cos Q_x - 2\cos Q_y$.

We have analyzed the above equations to extract information 
about the importance of quantum and thermal phase fluctuations.
Our numerical results can be simply summarized as follows:
$\la \delta\theta^2 \ra (T=0) \sim
\sqrt{(e^2/\epsilon_b\xi_0)/(D_{_\pll}(0) d_c)}$ is a measure of
quantum fluctuations, while
thermal fluctuations become important above a crossover scale
$T_\times \sim \sqrt{(D_{_\pll}(0)d_c)(e^2/\epsilon_b\xi_0)}$. 
These quantities are simply understood as the zero point spread and
the energy level spacing of an oscillator respectively, in the renormalized 
harmonic theory. The scale $T_\times$ is also the temperature at which 
$n_c(\vQ=\pi) \sim 1$ (giving a broad kernel $K(z)$), with non-local 
effects in $\tau$ becoming important. 

It is easy to see that phase fluctuation effects are negligible in the 
BCS limit of large $\xi_0$. 
With $e^2/\epsilon_b a \sim D_{_\pll} d_c \sim E_{_F}$
and $\xi_0\sim v_{F}/\Delta$, one obtains the standard result
$E_{\rm cond} \sim \Delta^2/E_{_F}$ per unit cell, and
$\la \delta\theta^2 \ra (T=0) \sim \sqrt{\Delta/E_{_F}} \ll 1$ and
$T_\times \sim \sqrt{E_{_F}\Delta} \gg T_c$.

For the cuprates the short coherence length and small $D_{_\pll}$ act
together to increase $\la \delta\theta^2\ra$, but they push $T_\times$
in opposite directions. For optimal Bi2212 we use $e^2/\epsilon_b a \approx 
0.3eV$ with $\epsilon_b \approx 10$ and 
$\xi_0/a \approx 10$. Bi2212 has a bilayer stacking structure with the
planes within a bilayer being much closer than the distance between
bilayers. Assuming the phase within the bilayer to be fully
correlated, we set $d_c$ to be the mean inter-{\em bilayer} spacing and thus 
$d_c/a \approx 4$. Using $\lambda_{_\pll}(0)\approx 2100 \AA$, we then get 
the bilayer stiffness $D_{_\pll}(0) d_c \approx 75 meV$ \cite{lambda}. These 
values lead to $E_{\rm cond} 
\approx 6 K/{\rm unit cell}$ which is somewhat larger than estimates for 
optimally doped YBCO from specific heat 
measurements \cite{loram90-94}; we are unaware of similar data for Bi2212. 
We find that the crossover scale $T_\times \approx 350K \gg T_c$. 
Since the bare stiffness $D^0_{_\pll}$ actually decreases
with temperature due to quasiparticle excitations, a better estimate of the 
thermal crossover scale may be obtained from $T_\times \sim 
\sqrt{D^0_{_\pll}(T_\times)(e^2/\epsilon_b \xi_0)}$; this results in a
crossover scale $T_\times \sim T_c$. Thus, thermal 
fluctuations are clearly unimportant at low temperatures $T\ll T_c$. Quantum 
fluctuations are important since we find $\la \delta\theta^2 \ra(T=0) 
\sim 1$ at optimal doping. 

\begin{center}
\begin{figure}
\leavevmode
\epsfig{figure=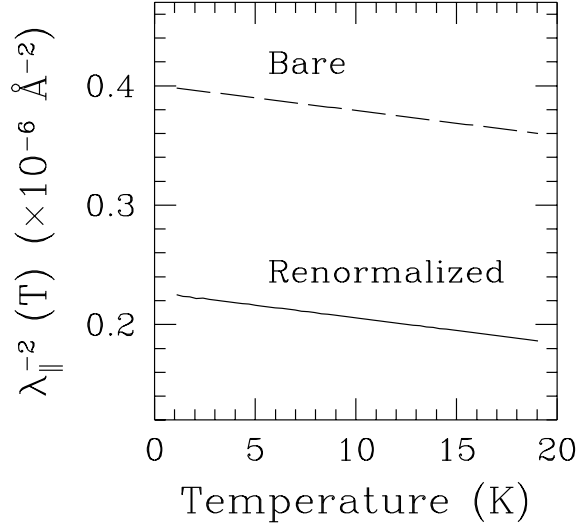,width=8cm,angle=0}
\caption{The bare and renormalized $1/\lambda^2_{_\pll}$ plotted as 
a function of temperature near optimal doping for Bi2212. We have chosen
bare values such that the renormalized results, $\lambda_{_\pll}(0)
\approx 2100\AA$ and $d\lambda_{_\pll}/dT\approx 10.0 \AA/K$, are in 
approximate agreement with experiment.
}
\label{schafig}
\end{figure}
\end{center}

To study the temperature dependence of $\lambda_{_\pll}(T)$ and the
bilayer stiffness $D_{_\pll}(T) d_c$ \cite{lambda}, we set the bare
stiffness $d_c D^0_{_\pll}(T) =d_c  D^0_{_\pll}(0)- 2 \alpha^0 T$, 
where the linear decrease arises purely from nodal quasiparticle excitations 
within a single layer. This implies
$1/\lambda_{_\pll,0}^2(T)=1/\lambda_{_\pll,0}^2(0)- \left(4\pi e^2/\hbar^2 c^2
d_c\right) 2 \alpha^0 T$.
We plot the results of a numerical calculation of $1/\lambda^2_{_\pll}(T)$ in 
Fig~(\ref{schafig}). Phase fluctuations are seen to lead to a large
quantum renormalization of $1/\lambda_{_\pll}^2(0)$ and to very little
change in the slope of $1/\lambda_{_\pll}^2(T)$ \cite{m2s00}. 
The negligible renormalization of the slope of $1/\lambda_{_\pll}^2(T)$ 
which we find, is true for a range of parameter values around our specific 
choice which has been constrained by experiments.
It is however not the case more generally, and the slope could be
renormalized by quantum fluctuations for a very different choice of 
parameter values. This effect of quantum fluctuations should be contrasted 
with the effect of {\em classical} thermal
phase fluctuations which do not renormalize $\lambda_{_\pll}(0)$ or
$D_{_\pll}(0)$, but increase the slope of $1/\lambda_{_\pll}^2(T)$ 
relative to its bare value.

We note that in the absence of quasiparticles, the superfluid stiffness
in this model would have an exponentially small temperature dependence,
arising from phase fluctuations which are gapped. 
While it might appear that there could
be low temperature crossovers resulting from the $c$-axis plasmon being at 
low energy ($\sim 10 K$ for Bi2212), the phase space for these low lying 
fluctuations is extremely small to lead to a linear $T$ behavior. Even in a 
purely 2D system, which supports (gapless) low energy plasmons dispersing as 
$\sqrt{q_{_\pll}}$, the phase stiffness
would decrease slowly, with a large power law ($\sim T^5$). Quasiparticles 
are thus crucial in obtaining the observed linear temperature dependence.

We find that we have to choose the bilayer stiffness 
$d_c D^0_{_\pll}(0)\approx 130 meV$
corresponding to a bare $\lambda_{_\pll,0}(0) \approx 1600 \AA$ and a 
slope
$\alpha^0 \approx 0.35 meV/K$ to obtain the renormalized values 
$d_c D_{_\pll} \approx 75 meV$ (implying $\lambda_{_\pll}\approx 2100 \AA$)
and $\alpha \approx 0.35 meV/K$, in agreement 
with experiment. Thus quantum effects lead to a large ($\sim 40\%$) decrease 
of $1/\lambda^2_{_\pll}(0)$, but no change in its linear $T$ slope 
\cite{m2s00} for our choice of parameter values.

The bare slope $\alpha^0$ within a theory 
of non-interacting Bogolubov quasiparticles is given by $(k_{_B}\ln 2 /\pi) 
(\hbar v_{F} k_{_F} /\Delta_d)$. Using the measured ARPES dispersion 
\cite{param}, this leads to $\alpha^0 \approx 0.8 meV/K$ which is much
larger than the 'bare' value we have used above to obtain agreement 
with penetration depth experiments. This points to the inadequacy of the 
non-interacting quasiparticle picture. The above discrepancy could be 
accounted for by considering quasiparticle interaction effects at the 
mean field level before considering the effect of phase fluctuations.
These interaction effects become more important as one underdopes to 
approach the Mott insulator\cite{mesot99}.

\section{conclusions} 

In this paper, we have focussed on the excitations of a short
coherence length $d$-wave superconductor. These are nodal
fermions and the fluctuations of the amplitude and phase of
the order parameter. Using an effective phase-only action 
we have discussed collective plasma modes and renormalization of
superfluid stiffness by anharmonic longitudinal phase fluctuations. We
summarize below some of our main conclusions.

We have found that the important
excitations are the low-lying fermionic states near the nodes and quantum
phase fluctuations of the order parameter. Although the $d$-wave state
supports in addition, two amplitude fields and a bond-phase field, these
have been shown to have negligible spectral weight at low energy and are
unimportant for the low temperature thermodynamics. They could possibly be
probed in experiments such as Raman spectroscopy measurements designed to 
detect these fluctuations.

Our discussion and derivation of the plasma modes emphasizes a unified way 
of looking at the in-plane and $c-$axis plasmons, in a manner which is
relatively independent of detailed models of $c-$axis propagation. The very 
different nature of the two plasma modes, with a small $c-$axis plasma 
frequency 
governed by the $c-$axis stiffness and a large in-plane plasma frequency not
directly related to the in-plane stiffness, can both be understood within
our phase action. A microscopic derivation of $c-$axis conductivity 
sum-rules and $T$-dependent spectral weight transfers would
depend on specific models \cite{dassarma98-1,ioffe99}, and we have not 
discussed these.

Our derivation and treatment of the effective phase-only action emphasizes
the crucial role played by the coherence length in imposing momentum and 
frequency cutoffs in the phase fluctuations. This allows us to interpolate 
from the
BCS limit where phase fluctuations are unimportant to a regime of strong
quantum fluctuations in the short coherence length limit. We find that
quantum and thermal fluctuations cannot both be present at low
temperatures; The short coherence length increases quantum fluctuations
while pushing up the temperature scale at which one crosses over to
thermal phase fluctuations. 
The strong longitudinal quantum fluctuations of the phase predicted 
by our calculation would also imply dynamical charge density fluctuations 
at low temperatures in the SC phase, which could possibly be probed in 
experiments. 

It has been pointed out that there is a discrepancy in the magnitude of
the linear $T$ slope of the measured penetration depth and the value
calculated using ARPES data assuming free quasiparticles \cite{mesot99}.
However, phase fluctuations effects had not been taken into 
account before comparing data from the two measurements.
We find that even including the relevant quantum phase fluctuations, 
a discrepancy is present which points to strong quasiparticle interactions
even at optimal doping. A theory to account for these quasiparticle 
interactions is however lacking. One possibility is to invoke a 
phenomenological superfluid Fermi liquid theory description for the 
quasiparticles\cite{millis98,m2s00}. It turns out however, that such a 
theory has a large number of free parameters and lacks predictive power 
although the experimental results may be easily rationalized.

Finally, we have restricted our study in this paper to the low temperature 
properties of the SC without addressing the issue of what happens at higher 
temperatures 
within the SC state and in the normal state. This leads naturally to the
problem of fermions interacting with a strongly fluctuating order
parameter, which is at present an important open problem.

\bigskip

{\em Note added in proof:} We have recently studied, in some detail, the 
effect of ohmic dissipation
on phase fluctuations \cite{benfatto00}. Such dissipation, arising from a 
finite low frequency optical conductivity, is seen to reduce the magnitude 
of quantum fluctuations and reduce our estimate for the thermal crossover 
scale. Nevertheless, we find that the crossover scale is still large,
so that our conclusion, about quasiparticles dominating the low temperature 
behavior of response functions, remains unchanged.

\bigskip

{\bf Acknowledgements:}  A.P. thanks D. Gaitonde and C. Panagopoulos 
for useful discussions. The work of M.R. was supported in part by the D.S.T.,
(Govt. of India) under the Swarnajayanti scheme.

\bigskip

\section*{Appendix A}

In this Appendix, we present an approximate analysis of the density of
states for the amplitude fields $\eta_{s,d}$ and the bond-phase field
$\phi$. This analysis gives us insight into the nature of fermionic 
excitations which contribute to the low energy spectral weight for these 
fields and recovers the power law for the low energy DOS obtained in the 
numerics.

\medskip

{\bf A.1 Amplitude fluctuations:}

\smallskip

The low energy density of states, $N_i (\omega) = \frac{1}{N}
\sum_\vq -{\cal I}m 
M_i(\vq,\omega)/\pi ~~~i=s,d$ with the restriction $|\vq_x|,|\vq_y| <
\pi/\xi_0$. The spectral weight ${\cal I}m M_i(\vq,\omega)$ at $T=0$
arises from summing over low lying pair excitations at 
{\em all} momenta $(\vk,\vk-\vq)$.
This contributes to absorption at frequencies $\omega=E_\vk+E_{\vk-\vq}$.
For $\vq=0$, the spectral weight extends to $\omega=0$ coming from low
lying pair excitations from momenta $\vk$ arbitrarily close to the node.
At finite $\vq$ the absorption sets in beyond a minimum threshold which 
corresponds to $\vk$ at the node and $\vk-\vq$ near the node 
(since $\vq$ is small due to the momentum cutoff). The threshold is given by
$\omega_{min}(\vq)=\sqrt{q_1^2 \Delta_d^2+q_2^2 v^2_{_F}}$ where $q_1,q_2$ 
refer to components 
of $\vq$ parallel and perpendicular to the Fermi surface respectively,
at the node.

We find numerically that the spectral weight for $\vq \neq 0$ is nearly
the same as for $\vq=0$ beyond the absorption threshold. We therefore
approximate
\be
N_i(\omega) \approx \left(\frac{-1}{\pi}\right)
\int_{-\pi/\xi_0}^{\pi/\xi_0} ~\frac{d^2\vq}{(2\pi)^2} 
{\cal I}m M_i(\vq=0,\omega) \Theta(\omega-\omega_{min}(\vq))
\ee
where $\Theta(x)$ is the unit step function which is $1$ for $x>0$ and
$0$ otherwise. For $\vq=0$, the inverse propagator
\be
M^{-1}_{s,d} = \frac{\Delta_d^2}{4 J} - \frac{\Delta_d^2}{2 N}\sum_\vk
\varphi^2_{s,d}(\vk) \frac{\xi_\vk^2}{E_\vk (\omega_n^2+4 E^2_\vk)}
\ee
where $\varphi_s(\vk)=\cos k_x + \cos k_y$ and
$\varphi_d(\vk)=\cos k_x - \cos k_y$.
Analytically continuing $i\omega_n \to \omega+i 0^{+}$ 
and working at low frequency ($\omega \ll \Delta_d$) leads to 
\be
-\frac{1}{\pi} {\cal I}m M_{s,d}= c_{s,d}^2 \frac{1}{N}
\sum_\vk \delta(E_\vk - \omega/2)~\varphi_{s,d}^2(\vk)~\xi^2_\vk/E^2_\vk
\label{Aeta}
\ee
where $2/c_{s,d} \equiv \Delta_d \frac{1}{N}\sum_\vk\left(\varphi^2_d/E_\vk - 
\varphi_{s,d}^2 \xi^2_\vk / E^3_\vk\right)$.
We evaluate the $\omega$-dependent momentum sum in (\ref{Aeta})
analytically by converting
it to a Fermi surface integral and compute the constant $c_{s,d}$ 
numerically. Finally, doing the $\vq$ sum to obtain the density of states 
leads to:
\bea
\frac{N_s(\omega)}{N_{qp}(\omega)}&=& c_s^2~
\frac{\varphi_s^2({\bf k}_{_F,n})}{16\pi} 
\frac{\omega^2}{\Delta_d v_{_F}}\\ \nonumber
\frac{N_d(\omega)}{N_{qp}(\omega)}&=& c_d^2~ 
\frac{1}{256\pi} \frac{\omega^4}{\Delta_d^3 v_{_F}}
\label{dosamp}
\eea
where $\varphi_s({\bf k}_{_F,n})$ refers to $\varphi_s$ evaluated at the 
gap node point on the Fermi surface and $N_{qp}(\omega)\equiv 
k_{_F}\omega/(\pi v_{_F}\Delta_d)$ 
is the quasiparticle DOS per spin which is linear in $\omega$. 
We numerically estimate $c_{s,d}\sim 10$; The prefactors in the 
(\ref{dosamp}) are then of order unity.

\noindent{\bf A.2 The ``bond-phase'' field $\phi$:}

For the $\phi-$field, pair excitations similar to that for amplitude
excitations lead to low energy spectral weight. This is easy to
understand since both fields couple to the particle-particle channel
with only different vertex factors. The behavior of ${\cal I}m
M_\phi(\vq,\omega)$ is similar to that for amplitude fields with 
a vanishing threshold for $\vq=0$ and a finite threshold for $\vq\neq
0$. Following similar approximations, we set
\be
N_\phi(\omega) =  -\frac{1}{\pi}
\int_{-\pi/\xi_0}^{\pi/\xi_0}
\frac{d^2\vq}{(2\pi)^2} {\cal I}m M_\phi(\vq=0,\omega) 
\Theta(\omega-\omega_{min}(\vq))
\ee

For $T=0$, using particle-hole symmetry near the Fermi surface
we find the inverse propagator 
\be
M_\phi^{-1}(\vq=0,\omega_n)
=\frac{\Delta^2_d}{16} \frac{1}{N}\sum_{\vk}
\left[ \frac{\varphi^2_d(\vk)}{E_\vk} - 8 
\cos^2(k_y) \frac{E_\vk}{\omega^2_n+4 E^2_\vk} \right]
\ee
Doing the integrals as before, we finally get
\be
N_\phi(\omega)/N_{qp}(\omega) 
= \frac{1}{8 \pi}~c_\phi^2~\varphi_s^2({\bf k}_{_F,n})
\frac{\omega^2}{\Delta_d v_{_F}}
\ee
where $1/c_\phi \equiv \frac{\Delta_d}{N}\sum_\vk \varphi^2_s(\vk)/2E_\vk$.
The prefactor here is again of order unity.

\section*{Appendix B}

In this Appendix, we briefly consider the 
linear time derivative term in the phase action, which
we have dropped in the paper. 
For notational
simplicity, we consider a neutral $s$-wave SC in 2D with lattice 
spacing $a=1$.
In carrying out the Hubbard Stratonovitch transformation as for the $d$-wave
case, we introduce complex order parameter fields
$\Delta_\vr(\tau), \Delta^*_\vr(\tau)$ which are bosonic variables and satisfy
the constraint $\Delta_\vr(0)=\Delta_\vr(1/T)$.
Writing $\Delta_\vr=|\Delta_\vr| e^{i\theta_\vr}$, this translates into
$|\Delta_\vr|(1/T)=|\Delta_\vr|(0)$, and
$\theta_\vr(1/T)=\theta_\vr(0)+2\pi m_\vr$,
and the partition function involves an additional sum over $m_\vr$.
Working in small $\theta$ gradients
and doing a cumulant expansion, one arrives 
at the following form of phase action for low momenta and frequencies:
\be
S_\theta = \int_0^{1/T}d\tau~\sum_\vr \left[ i \rho \dot{\theta} + 
\frac {1}{8}\kappa \dot{\theta}^2 + \frac{1}{8}D_s(\nabla\theta)^2\right]
\ee
We now make the substitution $\theta(\vr,\tau)=\Theta(\vr,\tau)+ 2 \pi T m_\vr
\tau$ which implies $\Theta(\vr,1/T)=\Theta(\vr,0)$. Substituting
this in the action, we get
\begin{eqnarray}
S&=&2\pi i \rho \sum_\vr m_\vr + \frac{1}{2}\pi^2 \kappa T \sum_\vr m^2_\vr + 
\frac{D_s \pi}{6 T} \sum_\vr (\nabla m_\vr)^2 + \frac{1}{2}\pi D_s T
\sum_\vr \int_0^{1/T} d\tau~~\tau (\nabla \Theta)\cdot(\nabla m_\vr) \\
&+&\frac{1}{8}\int_0^{1/T}\sum_\vr \left(\kappa \dot{\Theta}^2+D_s (\nabla
\Theta)^2 \right)
\end{eqnarray}
where the derivatives denote discrete derivatives on the lattice. 

Now, at very
low temperatures, $D_s/T \gg 1$, we must set $\nabla m_\vr =0$ while at
high temperatures, $\kappa T \gg 1$, we must set $m_\vr=0$. In either case,
the field $\Theta$ decouples from the field $m_\vr$ and we get a Gaussian
theory of phase fluctuations. 
The former condition ($D_s/T \gg 1$) is equivalent to the condition that
the spatial phase variation due to thermal effects is small; in particular, 
vortex configurations are unimportant. The latter condition ($\kappa T \gg 1$) 
is just that the system starts 
behaving classically; since the extension along the imaginary time axis is
$1/T$, there is essentially no dynamics if $1/T 
\to 0$ and $\kappa$ is finite.

In the presence of vortices, the core would described
by a region where the magnitude of the order parameter $|\Delta_\vr|$ decreases
to zero. Since $|\Delta_\vr|$ is a bosonic variable, the core would 
trace a closed loop in ``time'' $1/T$.
In this case, all the electrons which lie inside the loop undergo
a phase change of $2\pi$ each time the loop is traced while electrons 
outside the 
loop return to their original phase angle. This leads to a Berry phase
factor $i \pi \rho S(\Gamma)$ for the loop $\Gamma$ with area $S(\Gamma)$.
The effect of this Berry phase factor on vortex dynamics was pointed 
out by Ao and Thouless\cite{ao93}, 
except they obtained a coefficient of $\rho_s$ (the superfluid density) 
instead of $\rho$ (the total electron density). Their result is
special to a Galilean invariant systems at $T=0$, where $\rho_s = \rho$.
Our result has also been derived earlier by Gaitonde and Ramakrishnan
\cite{dattu97}.

In the paper, we consider only slow spatial variations
of the phase and work with just the periodic variable $\Theta$ which we 
refer 
to as $\theta$. We note that the linear time derivative term will not be 
important
in the critical regime around the finite temperature superconductor to 
normal metal
phase transition where dynamics is unimportant. It would however be important
near quantum critical points at $T=0$ \cite{fwgf89}.

\section*{Appendix C}

Physically relevant correlation functions should be gauge invariant.
The functional integral method leads to a very simple and elegant
way of demonstrating the role of phase fluctuations in restoring
gauge invariance. (This is, of course, well known from early work
of Anderson and others \cite{pwa59,kadanoff61}). 

To this end we introduce external
gauge potentials $({\bf A},A_0)$ in the Hamiltonian. 
This leads to
the following modifications in the action (\ref{fullaction}).
In the expression (\ref{lag0}) for ${\cal L}_0$
we replace $\mu$ by $\mu + A_0(\vr,\tau)$ and 
$\delta\theta/2$ by $\delta\theta/2 - A_{\vr,\vr'}(\tau)$
where
$\delta\theta \equiv [\theta(\vr,\tau)-\theta(\vr',\tau)]$.
Consider the gauge transformation
$A_0(\vr,\tau) \to  A_0(\vr,\tau) + i \partial_\tau
\alpha(\vr,\tau)$ and 
$A_{\vr\vr'}(\tau) \to A_{\vr\vr'}(\tau) -(\alpha(\vr',\tau)
- \alpha(\vr,\tau))$. Invariance under this gauge transformation
implies that the phase field $\theta$ must transform as
$\theta(\vr,\tau) \to \theta(\vr,\tau) + 2 \alpha(\vr,\tau)$
with $|\Delta|$, $\phi$ and the fermion fields unchanged. 
The correlation functions ${\chi_0}$ and ${\Lambda_0}^{\alpha\beta}$ 
calculated at the mean field level, setting $\theta \equiv 0$ and 
$\phi \equiv 0$ are {\it not} gauge invariant.
From the above discussion, it is clear that this problem can be solved by
allowing for $\theta$ fluctuations and integrating over these
(rather than freezing $\theta \equiv 0$).
We now proceed to do this at the Gaussian level, which is entirely
equivalent to the old RPA calculation. 

\begin{figure}
\leavevmode
\epsfig{figure=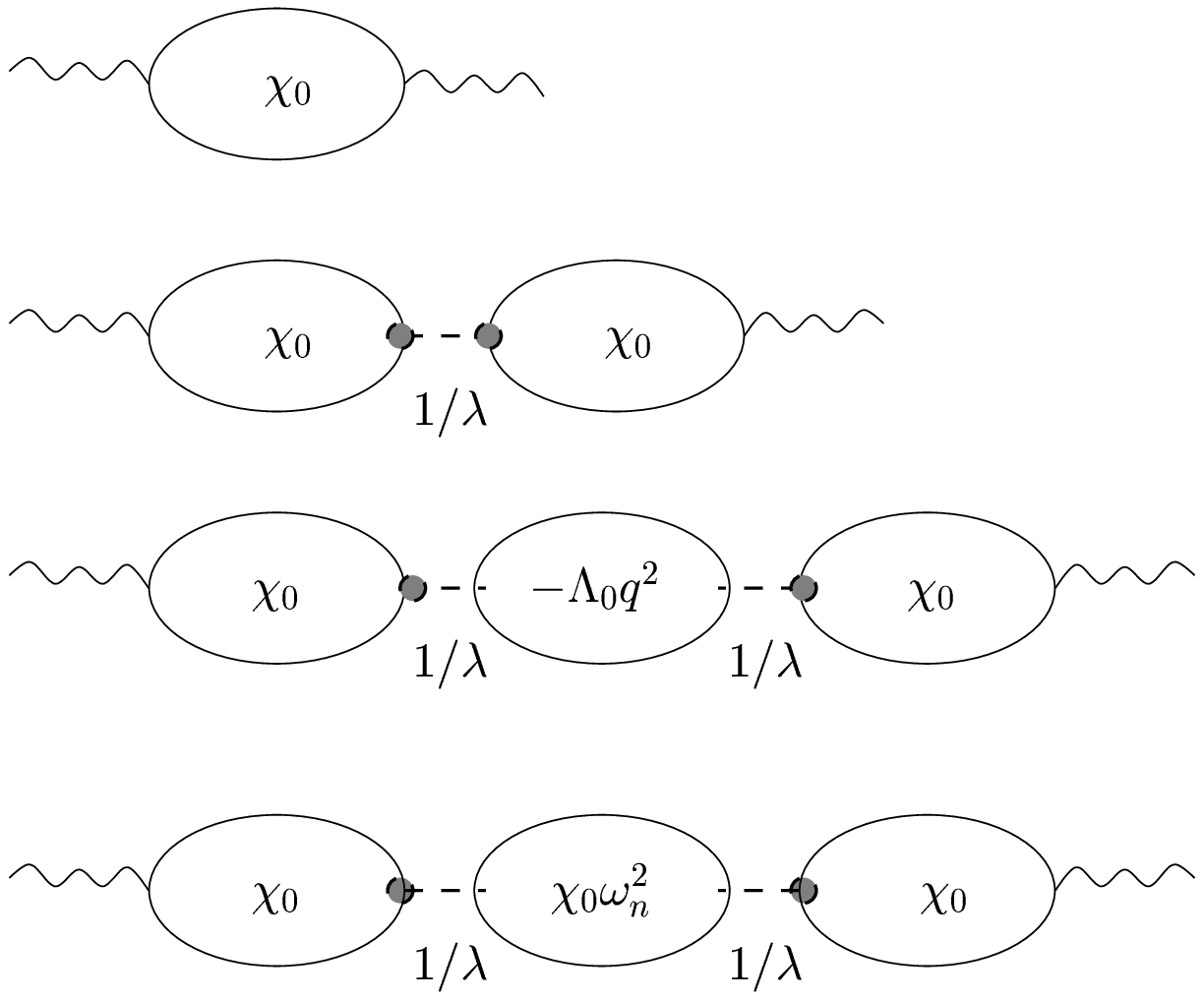,width=10cm,angle=0}
\caption{Low order diagrams for $\chi$: The wavy lines indicate $A_0$, 
the external
scalar potential. The dashed lines contribute $1/\lambda$, the ``fake'' 
$\theta$-propagator. The heavy dots refer to vertex factors of 
$\omega_n$ arising from the vertex $i\rho\partial_\tau\theta$. Finally,
the bubble contributions arising out of integrating out fermions are
explicitly indicated.
}
\label{feyn}
\end{figure}

Physical correlation functions are obtained by integrating out
the fermions and functional differentiation of the resulting
Gaussian effective 
action with respect to the external sources $A_0$ and ${\bf A}$. 
We emphasize that these sources couple minimally to the {\em original}
fermion operators, before the transformation 
$c^\dg \to c^\dg e^{-i\theta/2}$ in Section IV.
We then find the density-density correlation allowing for Gaussian phase
fluctuations to be
\be
\chi(\vq,i\omega_n)=
{{\vq^\alpha \vq^\beta {\Lambda_0}^{\alpha\beta} {\chi_0}}
\over{\vq^\alpha\vq^\beta {\Lambda_0}^{\alpha\beta} 
- \omega^2_n {\chi_0}}}.
\label{phyChi}
\ee
This result obtains diagrammatically as follows. The fermions couple
to the external source $A_0$ and to the phase field $\theta$ and we have
to integrate out both $\theta$ and the fermions. 
However $\theta$ itself does not have a propagator unless the fermions 
are integrated out. To simplify the diagrammatic calculation we first 
introduce a fake term $\lambda \theta(q)\theta(-q)$ in the action which 
leads to 
the bare $\theta$-propagator $1/\lambda$; we will take the limit 
$\lambda \to 
0$ at the end. The diagrams in Fig.\ref{feyn}
result for $\chi$. Summing the geometric series leads to
\be
\chi = {\chi_0} + \frac{{\chi_0}^2 \omega_n^2}{\lambda}
\Big[ 1- \left(\frac{{\chi_0}\omega_n^2-\Lambda_0 \vq^2}{\lambda}\right)
\Big]^{-1}
\ee
Taking the limit $\lambda\to 0$ then leads to the result in (\ref{phyChi}).

We note that the physical static compressibility is given
by $\chi(\vq\to 0,\omega_n=0)= {\chi_0}(\vq\to 0,
\omega_n=0)$,
i.e., the mean field result is unaffected by phase fluctuations
at the RPA level.

Similarly, denoting the physical $(\vq,\omega)$ dependent stiffness by
$\Lambda$, we get
\be
\Lambda^{\alpha\beta}= {\Lambda_0}^{\alpha\beta}
+ {{{\Lambda_0}^{\alpha\mu}{\Lambda_0}^{\nu\beta}\vq^\mu \vq^\nu} 
\over {\left[\omega^2_n{\chi_0} - 
{\Lambda_0}^{\mu\nu} \vq^\mu \vq^\nu\right]}}
\label{phyP}
\ee
We see that the transverse phase stiffness, (for instance along the 
$x-$direction), given by 
$\Lambda^{xx}(\omega_n=0, \vq_\perp \to 0,q_x=0)$ is unaffected
by the Gaussian phase fluctuations. 
However, the longitudinal part of the current correlation function is 
affected, and
\bea
{1 \over T} \la j_x j_x \ra (\omega_n=0,\vq_\perp=0,q_x\to 0)&= &
{1 \over T} \la j_x j_x \ra_0(\omega_n=0,\vq_\perp=0,q_x\to 0)+
{\Lambda_0}^{xx}(\omega_n=0,\vq_\perp=0,q_x\to 0)\\
&=&  \frac{1}{\Omega}\sum_\vk m^{-1}_{xx}(\vk) \la n_\vk \ra,
\eea
now satisfies the $f$-sum rule (which was violated at mean field
level). Thus gauge invariance is restored. 

The derivation remains unchanged in charged systems, with the only 
difference
being that we have to make the replacement 
${\chi_0} \to \chi_0^{\small RPA} \equiv {\chi_0}/ (1-V_\vq{\chi_0})$
in all the equations (in this Appendix). This can be easily understood
by comparing (\ref{neutralaction}) and (\ref{chargedaction}) in the 
text. It is not
hard to show that the longitudinal conductivity ($\sigma_L$) defined through
the longitudinal dielectric function by 
\be
\epsilon\equiv\frac{1}{1+V_\vq \chi} = 1 + 
\frac{4\pi i \sigma_L}{\omega}
\label{epsilon}
\ee
in a gauge invariant theory,
and the transverse conductivity defined by $\sigma_T(\omega)=i \Lambda
e^2/(\epsilon_b\omega)$ (with $\Lambda$ being the transverse part of
$\Lambda^{\alpha\beta}$) are equal in the 
limit $\vq\to 0$. In the text, we
omit subscripts and refer to both conductivities by $\sigma$ since we work at
$\vq\to 0$. It is easy to see that $\sigma_T(\omega)$ and hence
$\sigma(\omega)$ is unaffected by phase fluctuations within RPA.
However, it is only in a gauge invariant theory, such as the RPA, that
(\ref{epsilon}) holds since it obtains from using the current 
conservation equation.

\vfill\eject

\vfill\eject

\begin{thebibliography}{999}
\bibitem{lambda}
In the London limit, we define $D_s$ to be related to the penetration depth 
$\lambda$ in a 3D bulk system through: $1/\lambda^2=4\pi e^2 D_s/\hbar^2 c^2$. 
The in-plane layer stiffness in an anisotropic layered system (with 
interlayer spacing $d_c$) is $d_c D_s$, and this has the dimension of 
energy.
\bibitem{param}
For various calculations, we have used a dispersion 
$\epsilon_\vk=-t_1(\cos k_x + \cos k_y)+t_2\cos k_x
\cos k_y$ with $t_1=0.3 eV, t_2= 0.16 eV$, a chemical potential $-0.15 eV$ 
corresponding to a hole doping of approximately $0.18$, and a SC gap 
$\Delta_d=35 meV$. These parameters are chosen to be in approximate 
agreement with photoemission data for optimal Bi2212. See 
ref.~\cite{arpes95}.
\bibitem{landausing}
At finite temperatures, the 
coefficients of the phase action at finite temperatures have Landau 
singularities at small $(\vq,\omega)$ coming from the quasiparticle 
contribution preventing a simple analysis.
\bibitem{perturb}
The quantum XY action can be derived perturbatively from a 
microscopic fermionic Hamiltonian in cases where one has a small parameter.
Examples include: Josephson junction arrays \cite{ambegaokar_NATO}
and the Bose limit of tightly bound Cooper pairs \cite{dePalo99}.
In both cases one exploits the small hopping amplitude of the electrons
to control the calculation.
\end{thebibliography}
\end{document}